\documentclass[final,12pt,twocolumn]{elsarticle}
\usepackage{amsmath,amssymb,wasysym}
\usepackage{graphicx}
\usepackage{color}
\usepackage{textcomp}
\usepackage{multirow}

\usepackage{after page} 
\usepackage{etoolbox}
\patchcmd{\MaketitleBox}{    \hrule\vskip12pt
    \ifvoid\absbox\else\unvbox\absbox\par\vskip10pt\fi
    \ifvoid\keybox\else\unvbox\keybox\par\vskip10pt\fi}{}{}{}
\apptocmd{\maketitle}{
\let\columnwidth=\textwidth
\hrule\vskip12pt
\ifvoid\absbox\else\unvbox\absbox\par\vskip10pt\fi
\ifvoid\keybox\else\unvbox\keybox\par\vskip10pt\fi
\hrule
\afterpage{\newpage{}\vspace*{14\baselineskip}}
}{}{}

\usepackage[switch]{lineno}
\usepackage[colorlinks=false,hidelinks]{hyperref} 
\usepackage{caption} 

\biboptions{sort,compress}

\journal{Int. J. Mass Spec.}
\begin{document}

\begin{frontmatter}

\title{An RF-only ion-funnel for extraction from high-pressure gases}

\author[SU]{T.~Brunner\corref{cor1}}
\ead{tbrunner@stanford.edu}
\cortext[cor1]{Corresponding author}
\author[SU]{D.~Fudenberg}
\author[ITEP,FAIR]{V.~Varentsov}
\author[SU]{A.~Sabourov\fnref{fn1}}
\fntext[fn1]{Now at Patrick AFB, FL, USA}
\author[SU]{G.~Gratta}
\author[TRIUMF]{J.~Dilling}
\author[SU]{R.~DeVoe}
\author[TRIUMF,CARLETON]{D.~Sinclair}
\author[CSU]{W.~Fairbank Jr.}
\author[UIB]{J.B.~Albert}
\author[ALA]{D.J.~Auty}
\author[DUKE]{P.S.~Barbeau}
\author[UIUC]{D.~Beck}
\author[CSU]{C.~Benitez-Medina}
\author[SLAC]{M.~Breidenbach}
\author[IHEP]{G.F.~Cao}
\author[CSU]{C.~Chambers}
\author[LAUR]{B.~Cleveland}
\author[UIUC]{M.~Coon}
\author[CSU]{A.~Craycraft}
\author[SLAC]{T.~Daniels}
\author[UIB]{S.J.~Daugherty}
\author[ALA]{T.~Didberidze}
\author[DREXEL]{M.J.~Dolinski}
\author[CARLETON]{M.~Dunford}
\author[OAK]{L.~Fabris}
\author[LAUR]{J.~Farine}
\author[MUNCH]{W.~Feldmeier}
\author[MUNCH]{P.~Fierlinger}
\author[LHEP]{R.~Gornea}
\author[CARLETON]{K.~Graham}
\author[LLNL]{M.~Heffner}
\author[ALA]{M.~Hughes}
\author[SU]{M.~Jewell}
\author[IHEP]{X.S.~Jiang}
\author[UIB]{T.N.~Johnson}
\author[UMASS]{S.~Johnston}
\author[ITEP]{A.~Karelin}
\author[UIB]{L.J.~Kaufman}
\author[CARLETON]{R.~Killick}
\author[CARLETON]{T.~Koffas}
\author[SU]{S.~Kravitz}
\author[TRIUMF]{R.~Kr\"ucken}
\author[ITEP]{A.~Kuchenkov}
\author[SB]{K.S.~Kumar}
\author[SEOUL]{D.S.~Leonard}
\author[CARLETON]{F.~Leonard}
\author[CARLETON]{C.~Licciardi}
\author[DREXEL]{Y.H.~Lin}
\author[UIUC]{J.~Ling}
\author[SD]{R.~MacLellan}
\author[MUNCH]{M.G.~Marino}
\author[LAUR]{B.~Mong}
\author[SU]{D.~Moore}
\author[SLAC]{A.~Odian}
\author[SU]{I.~Ostrovskiy}
\author[CARLETON]{C.~Ouellet}
\author[ALA]{A.~Piepke}
\author[UMASS]{A.~Pocar}
\author[TRIUMF]{F.~Retiere}
\author[SLAC]{P.C.~Rowson}
\author[CARLETON]{M.P.~Rozo}
\author[SU]{A.~Schubert}
\author[DREXEL]{E.~Smith}
\author[ITEP]{V.~Stekhanov}
\author[UIUC]{M.~Tarka}
\author[LHEP]{T.~Tolba}
\author[SU]{D.~Tosi}
\fntext[fn2]{Now at University of Wisconsin, Madison, WI, USA}
\author[SU]{K.~Twelker}
\author[LHEP]{J.-L.~Vuilleumier}
\author[UIUC]{J.~Walton}
\author[CSU]{T.~Walton}
\author[SU]{M.~Weber}
\author[IHEP]{L.J.~Wen}
\author[LAUR]{U.~Wichoski}
\author[UIUC]{L.~Yang}
\author[DREXEL]{Y.-R.~Yen}

\address[SU]{Physics Department, Stanford University, Stanford CA, USA}
\address[ITEP]{Institute for Theoretical and Experimental Physics, Moscow, Russia}
\address[FAIR]{Facility for Antiproton and Ion Research in Europe (FAIR GmbH), Darmstadt, Germany}
\address[TRIUMF]{TRIUMF, Vancouver BC, Canada}
\address[CARLETON]{Physics Department, Carleton University, Ottawa ON, Canada}
\address[CSU]{Physics Department, Colorado State University, Fort Collins CO, USA}
\address[UIB]{Physics Department and CEEM, Indiana University, Bloomington IN, USA}
\address[ALA]{Department of Physics and Astronomy, University of Alabama, Tuscaloosa AL, USA}
\address[DUKE]{Department of Physics, Duke University, and Triangle Universities Nuclear Laboratory (TUNL), Durham North Carolina, USA}
\address[UIUC]{Physics Department, University of Illinois, Urbana-Champaign IL, USA}
\address[SLAC]{SLAC National Accelerator Laboratory, Menlo Park CA, USA}
\address[IHEP]{Institute of High Energy Physics, Beijing, China}
\address[LAUR]{Department of Physics, Laurentian University, Sudbury ON, Canada}
\address[DREXEL]{Department of Physics, Drexel University, Philadelphia PA, USA}
\address[OAK]{Oak Ridge National Laboratory, Oak Ridge TN, USA}
\address[MUNCH]{Technische Universitat Munchen, Physikdepartment and Excellence Cluster Universe, Garching, Germany}
\address[LHEP]{LHEP, Albert Einstein Center, University of Bern, Bern, Switzerland}
\address[LLNL]{Lawrence Livermore National Laboratory, Livermore CA, USA}
\address[UMASS]{Amherst Center for Fundamental Interactions and Physics Department, University of Massachusetts, Amherst MA, USA}
\address[SB]{Department of Physics and Astronomy, Stony Brook University, SUNY, Stony Brook NY, USA}
\address[SEOUL]{Department of Physics, University of Seoul, Seoul, Korea}
\address[SD]{Physics Department, University of South Dakota, Vermillion SD, USD}

\begin{abstract}
An RF ion-funnel technique has been developed to extract ions from a high-pressure (10\,bar) noble-gas environment into a vacuum ($10^{-6}$\,mbar).
Detailed simulations have been performed and a prototype has been developed for the purpose of extracting $^{136}$Ba ions from Xe gas with high efficiency.
With this prototype, ions have been extracted for the first time from high-pressure xenon gas and argon gas.
Systematic studies have been carried out and compared to simulations.
This demonstration of extraction of ions, with mass comparable to that of the gas generating the high-pressure, has applications to Ba tagging from a Xe-gas time-projection chamber for double-beta decay, as well as to the general problem of recovering trace amounts of an ionized element in a heavy (m\,$>40$\,u) carrier gas.
\end{abstract}


\end{frontmatter}
\section{Introduction} 
Ion extraction from gas environments at pressures $\gtrsim$1\,mbar is challenging since collisions dominate ion motion \cite{Kelly2010}.
Conventional mass-spectrometry techniques use a combination of skimmers and orifices to reduce the gas flow across differential pumping stages.
Similar ion-extraction techniques are applied in ion source assemblies at radioactive ion beam facilities, e.g. IGISOL LIST \cite{Reponen2011}.
Such techniques achieve low downstream pressures at some cost of efficient ion transport \cite{Kelly2010,Page2006,Page2007}.

Radio-frequency (RF) ion funnels have been developed for ion extraction with increased ion-transmission efficiency; typically, RF funnels extract from air into vacuum and improve ion extraction efficiency by more than an order of magnitude (see \cite{Ibrahim2006,Kelly2010} and references therein).
RF funnels have also been developed for ion transport in gas stopper cells \cite{Wada2013}.
Technical improvements have included blocking the gas jet with a jet disrupter electrode \cite{Kim2001} and a DC ion carpet at the exit \cite{Anthony2014}.
Various funnel designs have been realized, including a recent design using electrodes etched on PCBs \cite{Ibrahim2014}.
Typically, RF funnels are used with electrospray ion sources where gas is injected through a capillary.
Capillary inlet pressures reach an atmosphere and pressures inside the funnel reach $\sim$40\,mbar.
However, owing to the nature of this ion source, such funnels are not optimized for single ion extraction at close to 100\% efficiency and require a longitudinal DC potential gradient to transport ions through the funnel, adding complexity and components that can contaminate the vacuum.
To overcome these limitations and extend the use of RF-funnels to heavy-mass gases at a high-pressure, an RF-only ion funnel prototype has been developed. It aims to investigate the feasibility of extracting ions from 10\,bar Xe gas into a 1\,mbar vacuum in only one stage with high efficiency.
This prototype realizes a novel concept of ion transport via carrier gas flow instead of via applied DC-drag potentials that was suggested \cite{Varentsov2001} and described in detail \cite{Varentsov2002,Varentsov2004}.

The realization and trial of this RF-only funnel is an important step towards its application in the search for neutrinoless ($0\nu$) double-beta ($\beta\beta$) decay in $^{136}$Xe.
Over the past decades, experiments searching for the lepton-number non-conserving $0\nu\beta\beta$-decay \cite{Avignone2008} have excluded half lives (at 90\% CL) shorter than $T_{1/2}^{0\nu\beta\beta}\gtrsim10^{25}\,$yr (in $^{136}$Xe \cite{Auger2012,Gando2013,Albert2014} and in $^{76}$Ge \cite{Agostini2013}).
In order to extend experimental sensitivity, the development of larger detectors with reduced background is required.
For the detection of the $0\nu\beta\beta$ decay signal such a detector would ideally have no backgrounds.
The identification (Ba-tagging) of the atomic species produced in the decay, $^{136}$Ba for $\beta\beta$-decay of $^{136}$Xe, would drastically reduce the backgrounds that are dominated by radioactivity unrelated to the production of Ba in the detector.
The association of decay energy and event topology to the Ba-tagging technique \cite{Moe91} would allow for the discrimination between the 2-neutrino decay, which produces a continuum spectrum, and the interesting neutrinoless decay, which produces a mono-energetic line at the sum energy of the two electrons.

Among the $\beta\beta$-decay candidates, the possibility of tagging the final atomic state appears to be possible only for the case of $^{136}$Xe \cite{Moe91}.
The nEXO collaboration is developing a multi-ton $\beta\beta$-decay experiment using liquid $^{136}$Xe (LXe);
Ba-tagging technology appropriate for LXe is being developed \cite{Twelker2014,Mong2014} for possible use in a second phase of detector operation with sensitivity into the region of the normal hierarchy of neutrino mass \cite{Bilenky2012}.
A multi-ton detector using gaseous xenon (GXe) may be appropriate at a later stage to investigate the physics of $0\nu\beta\beta$-decay, should it be discovered by the LXe detector;
the relatively low-density gas would allow visualizing the tracks of low energy final state electrons.
Such a detector calls for the development of Ba-tagging for gas phase operations \cite{Danilov2000}.

Schematically, tagging of the Ba$^{++}$ daughter from gas xenon will be implemented in the following consecutive steps:
(i) The energy deposited and topology of each event is measured to determine whether it has a $\beta\beta$-like signature.
(ii) If it does, the electric fields inside the time-projection chamber are modified such that ions from the previously determined decay volume are drifted to an appropriate extraction port where they are flushed out by the Xe gas.
(iii) Ions are separated from the Xe gas and guided into vacuum.
(iv) Ba$^{++}$ is converted to Ba$^+$ by electron exchange (e.g. in triethylethane  gas \cite{Sin11}).
(v) The ion is captured in a linear Paul trap and is unambiguously identified by means of laser spectroscopy \cite{Green2007}.
The main challenge of this Ba-tagging method is the extraction of Ba-ions from a high-pressure Xe environment with near 100\% efficiency.

This paper describes the RF-only funnel apparatus built to investigate the extraction of Ba ions from xenon gas for application in Ba tagging and its tests.
The new technique is optimized for highly efficient extraction of single ions from an equal-mass carrier gas.
\begin{figure*}
\centering
		\includegraphics[width=\textwidth]{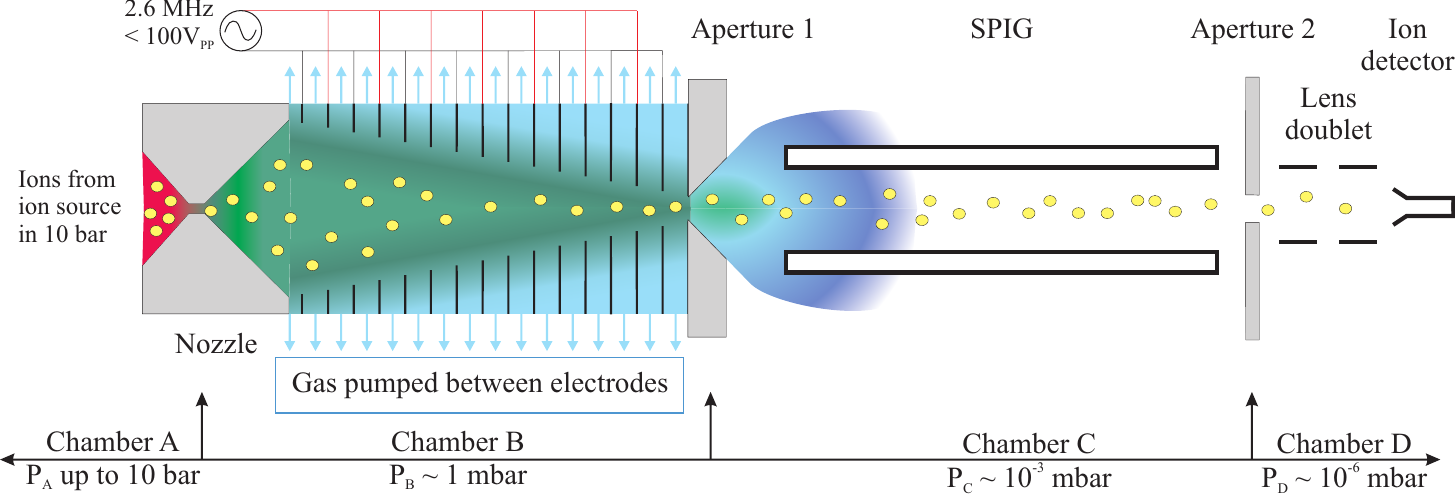}
		\caption{(color online)
Concept of the RF-only funnel prototype.
See text for explanation.%
}
	\label{fig:funnel-concept}
\end{figure*}

\section{Apparatus}
A schematic of the system is shown in Figure\,\ref{fig:funnel-concept}.
It has been optimized in gas dynamic calculations described in Section\,\ref{sec:simulations}.
Ions are created in a high-pressure noble-gas environment at the entrance of a converging-diverging supersonic nozzle.
The ions are injected through this nozzle via the supersonic gas flow into the conical cavity of the RF-funnel.
RF-voltage confines the ions while the majority of gas escapes and is pumped by a high capacity cryopump.
Exiting the RF-funnel $\sim$0.5\,ms later, the ions cross into a second differential pumping stage where they are captured by a sextupole ion guide (SPIG).
This guide transports the ions to a downstream chamber for detection, currently by a channel electron multiplier (CEM).
Details of the individual subsystems follow.
\subsection{Vacuum and gas handling system}\label{sec:VacSys}  
\begin{figure*}
	\centering
	\includegraphics[width=\textwidth]{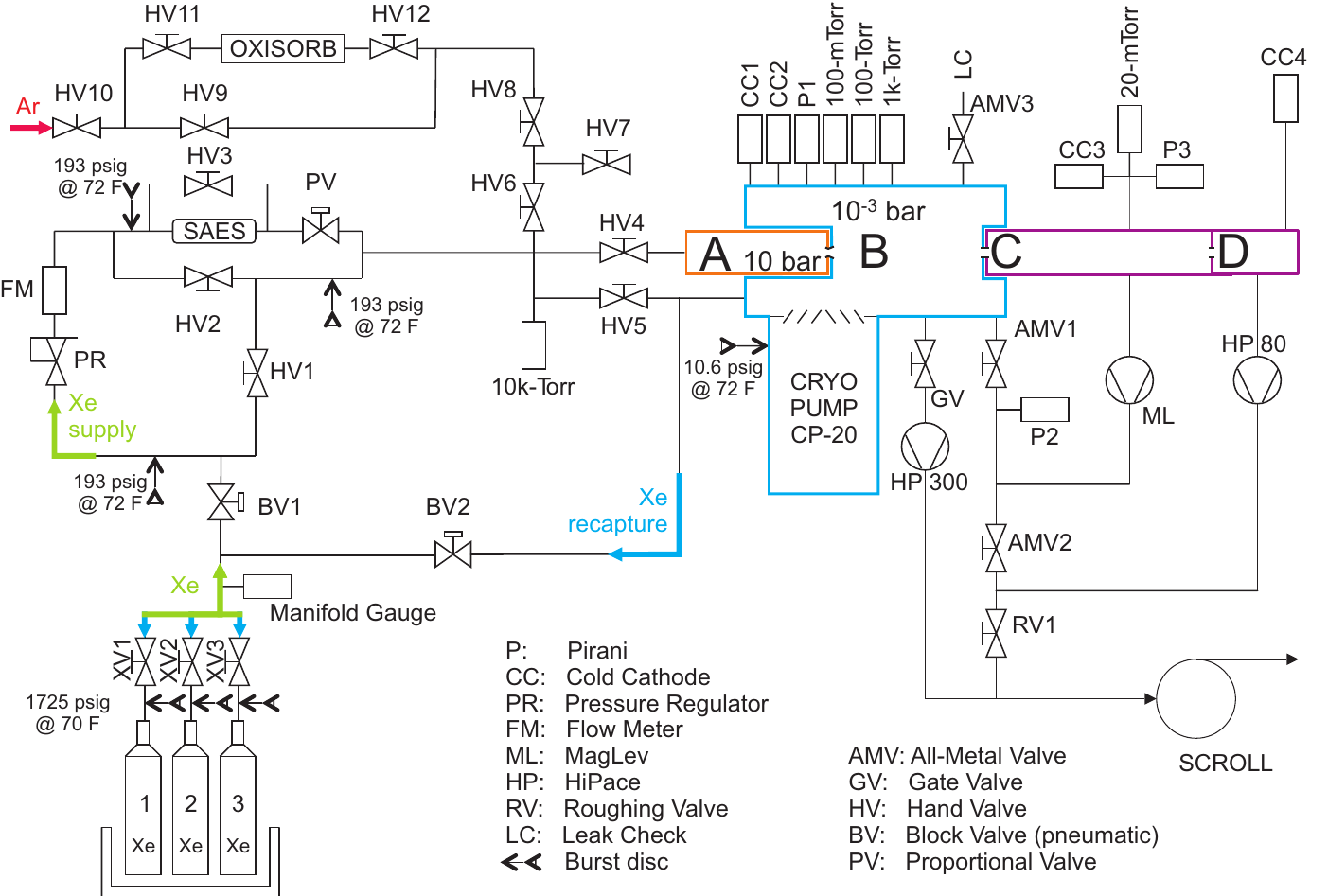}
	\caption{(color online)
Vacuum and gas system schematic.
The vacuum system consists of four chambers:
(A) high-pressure chamber,
(B) cryopump chamber with the RF-funnel installed at its center,
(C) SPIG chamber and
(D) detection chamber.
The chambers are separated by small apertures or the nozzle.
Arrows depict the flow direction of the gas.
Either argon (red arrow) or xenon (green arrows) gas can be supplied to chamber A.
Xenon is recaptured (blue arrows).
The baratron gauges are labeled by range.
See \hyperref[sec:VacSys]{text} for details.%
}
	\label{fig:pumping-scheme}
\end{figure*}

The vacuum and gas handling systems are designed to ultra-high vacuum (UHV) standards.
Only UHV compatible materials (with the exception of O-rings at apertures, in vacuum pumps and along forelines) were used in the funnel and the system, and trapped volumes were vented to avoid virtual leaks.
Great effort is taken to avoid contamination of the system that could affect ion extraction or transport.
Each part was ultrasonically cleaned for $\gtrsim$15\,minutes each in acetone and then ethanol.
A schematic of the gas handling and vacuum setup is shown in Figure\,\ref{fig:pumping-scheme}.
The vacuum system has four chambers:
(A) high-pressure chamber with ion source installed at the nozzle entrance,
(B) cryopump chamber with RF-funnel installed at center,
(C) SPIG chamber and
(D) detection chamber.
The converging-diverging nozzle separates chambers A and B, a $\diameter 1$\,mm aperture separates chambers B and C, and a $\diameter 2$\,mm aperture separates chambers C and D;
these act as differential pumping barriers.
The aperture between chambers C and D can be biased whereas the others are at ground potential.

During pump down, chambers A and B are evacuated by a turbo-molecular pump (TMP)\footnote{\label{foot:HP300}Pfeiffer HiPace300}, chamber C by a magnetically levitated (ML) TMP\footnote{Edwards STP-A803C}, and chamber D by a third TMP\footnote{Pfeiffer HiPace80}.
All TMPs are backed by the same scroll pump\footnote{Edwards XDS35i}.
The system's base pressures achieved with only the TMPs, are $1.1\cdot 10^{-8}$\,mbar, $2.3\cdot 10^{-9}$\,mbar, and $3.9\cdot 10^{-9}$\,mbar in chambers B, C, and D, respectively.
These base pressures decrease to $7\cdot 10^{-10}$\,mbar, $2.0\cdot 10^{-9}$\,mbar, and $3.1\cdot 10^{-9}$\,mbar in chambers B, C, and D, respectively, when the cryopump\footnote{Sumitomo Marathon CP-20 Cryopump} is activated.
The TMP on chamber B is only used to evacuate the system; during gas-jet operation, a gate valve (GV in Fig.\,\ref{fig:pumping-scheme}) separates it from the system.

During Xe operations and after pump down, the bulk of the Xe gas is pumped by the large cryopump in chamber B so that the gas can be recycled and kept clean.
The setup can also be operated with Ar gas.
Various stagnation pressures in chamber A, referred to as P$_{\textnormal{A}}$, have been used.

The system is operated in either recovery or non-recovery mode.
In recovery mode, all gas injected is captured by the cryopump in chamber B, which also backs the TMPs on chambers C and D.
This is achieved by closing valves GV and RV1 and opening valves AMV1 and AMV2 in Figure\,\ref{fig:pumping-scheme}.
In this mode the background pressure in chamber B, P$_{\textnormal{B}}$, is limited to protect the TMPs.
In non-recovery operation, the scroll pump is coupled to the TMPs on chambers C and D by opening RV1 and AMV2 while closing AMV1 and GV.
In this mode of operation the pressure in the foreline of the turbo pumps is independent of P$_{\textnormal{B}}$ thus allowing higher P$_{\textnormal{B}}$.
All gas entering chambers C and D is exhausted to atmosphere.
For reasons of cost-saving, recovery mode is always used for Xe.
Typically, non-recovery mode is chosen for Ar.

Xenon gas is supplied from three stainless steel cylinders and is purified by a SAES\footnote{SAES MC400-903:\hspace*{10pt}Acids, Bases  $<$ 1\,ppbV\\\hspace*{17pt}Organics, Refractory Compounds $<$ 1\,ppbV\\\hspace*{54pt} O$_2$, H$_2$O, CO, CO$_2$, H$_2$ $<$ 10\,pptV} getter.
BV1, PV, and HV4 are open to operate with a xenon jet.
To recover the xenon accumulated in the cryopump, the xenon storage cylinders are cooled by immersion in LN$_2$ and then the pump is warmed up with BV1 closed and BV2 open.
A final $\lesssim$10\,g of Xe cannot be recovered from the vacuum system.

Operation with an Ar gas jet is more convenient since not recovering the gas allows for faster turnaround time between tests.
Argon is supplied from a cylinder\footnote{Praxair AR 6.0RS 99.9999\% pure}.
Tests have been performed where the Ar was passed through an {OXI}SORB\footnote{OXISORB: O$_2<$ 5\,ppb, H$_2$O $<$ 30 ppb} purifier before  flowing into chamber A.
However, the studies found that this purifier did not improve ion extraction stability; thus all count-rate measurements with argon presented in this manuscript were performed using gas supplied directly from the cylinder.
The Ar supply is indicated by a red arrow in Figure\,\ref{fig:pumping-scheme}.

Burst discs\footnote{Continental Disc Corporation: 193\,psig @ 72$^{\circ}$F}  protect the gas lines;
this limits the maximum pressure in chamber A to 14.3\,bar.
The maximum operating pressure used in chamber A is 10\,bar.
The vacuum system is protected by a 1.7\,bar burst disc\footnote{MDC BDA-275-ASME:
10.6 psig @ 72$^{\circ}$F}.
\subsection{Ion source and gas nozzle}\label{sec:ionsource} 
A custom built $^{148}$Gd-driven Ba-ion source, similar to the one in \cite{Diez2010} is used to produce Ba ions.
The source is made by electroplating $^{148}$Gd onto a stainless steel plate (9.5\,mm\,$\times$\,6.4\,mm\,$\times$\,0.5\,mm) for a total activity of $\sim$144\,Bq.
A layer of 20\,nm of BaF$_2$ was evaporated over the $^{148}$Gd.
The Sm nucleus recoiling from the $^{148}$Gd $\alpha$ decay ($Q=3182.69$\,keV \cite{NuDat}) knocks out Ba or BaF ions from this layer.
Based on $\alpha$ rate and \cite{Diez2010} a Ba$^{+}$ rate of $\sim$59\,Hz is calculated. 

\begin{figure}
	\centering
	\includegraphics[width=\columnwidth]{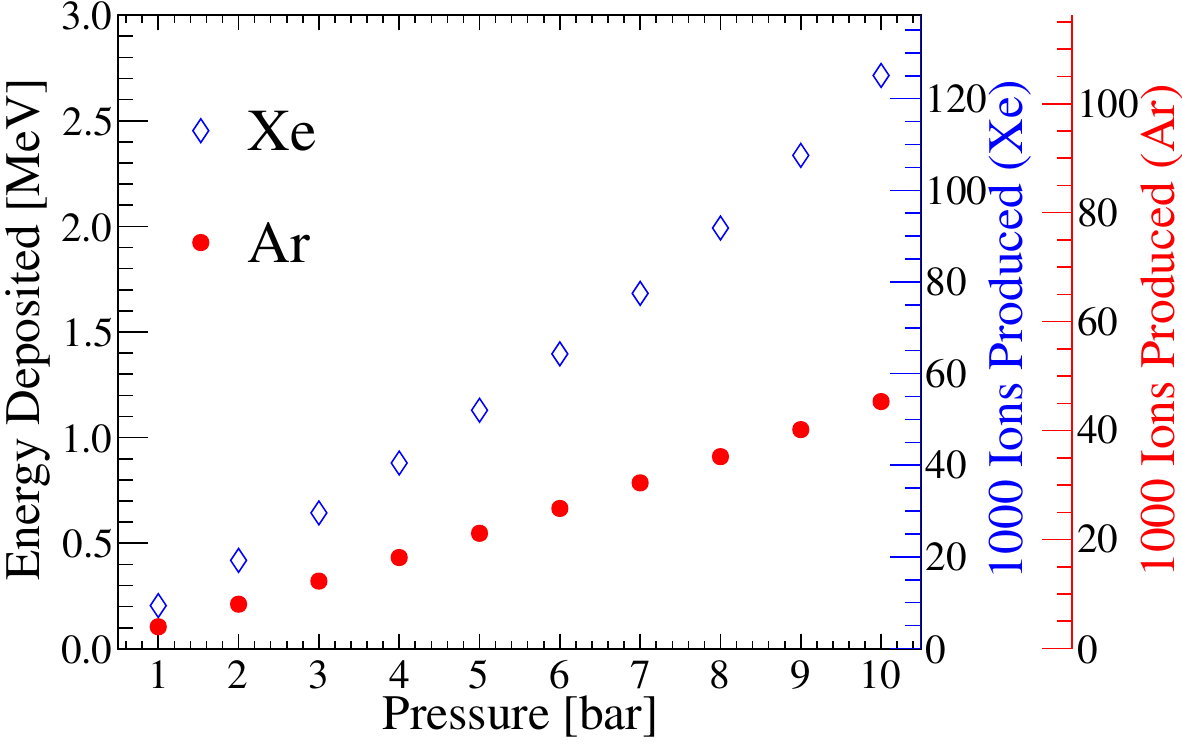}
	\caption{(color online) 
Simulated energy deposition through ionization by a 3183\,keV $\alpha$-particle in 0.91\,mm for xenon and argon. 
On the right hand side, separate axes for xenon and argon show the number of estimated ion-electron pairs produced.%
}
	\label{fig:ion-production}
\end{figure}

This source is installed in the carrier gas flow.
Alpha-particles from the source ionize the gas, so that ions other than $^{136}$Ba are expected to be produced.
The energy deposited along a 0.91\,mm travel distance in gas at 20\,$^{\circ}$C by such an $\alpha$-particle was calculated in SRIM2013 \cite{Ziegler2010} for various pressures of xenon and argon and is shown in Figure\,\ref{fig:ion-production}.
The calculated number of ion-electron pairs produced is also shown (right axis) \mbox{using} (the mean energy required to produce an electron-ion pair) W=21.7\,eV for xenon and W=25.8\,eV for argon \cite{Borges1996}. 
An experimental verification of these results is not possible for a number of reasons:
1) The source geometry is complex; 
for an $\alpha$-particle traveling perpendicular to the gas flow, the minimum distance through the gas is 0.48\,mm, the maximum distance is 2.1\,mm, and the average distance is 0.91\,mm.
2) It is not known how many of the electron-ion pairs recombine nor the pressure dependance.
3) The ratio of electron-ion pairs created in the gas versus Ba ions knocked out of the BaF$_2$ depends on pressure. 
\begin{figure}
	\centering
		\includegraphics[width=\columnwidth]{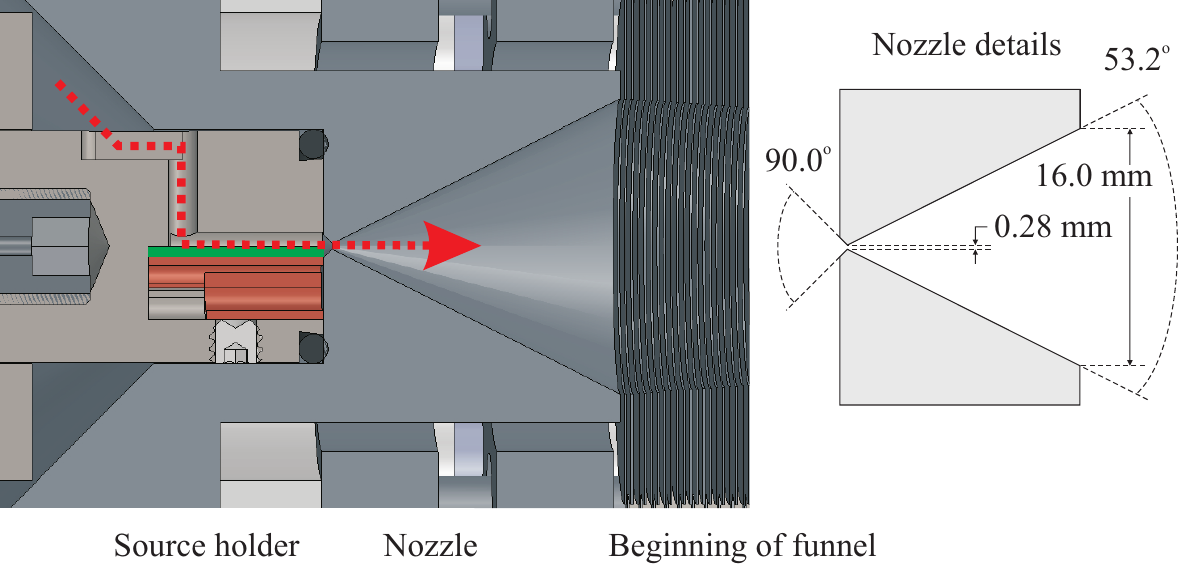}
	\caption{(color online) 
(left) Section view of source holder and (right) nozzle details.
Gas (red, dashed arrow) flows across the surface of the source plate (green line). 
Ions from the ion source are injected into the RF-funnel through the nozzle.
The start of the RF-funnel is shown at the exit of the nozzle.%
}
	\label{fig:Source-holder}
\end{figure}

A schematic view of the nozzle region with the source installed is shown in Figure\,\ref{fig:Source-holder}. 
A holder fixes the position of the source plate at the entrance to the nozzle in chamber A. 
An O-ring seals the holder to the nozzle, forcing gas through the holder and parallel to the surface of the source plate.
The atoms and ions created by the source are thus carried by the gas flow and injected into the funnel through the nozzle. 
The converging-diverging nozzle has a subsonic half-angle of 45$^{\circ}$ and a supersonic half-angle of 26.6$^{\circ}$. 
These two cones were machined using electric discharge machining\footnote{EDM Labs Ltd., www.edmlabsltd.com} in a DN40 (CF2.75$''$) flange.
The design value for the throat diameter was 0.30\,mm, however, the machined diameter is only 0.28\,mm. 
The design values for subsonic and supersonic part lengths are 0.5\,mm and 15.7\,mm, respectively, and the exit diameter of the supersonic nozzle is 16.0\,mm.
\subsection{RF-only funnel}\label{sec:funnel} 
In an RF-only funnel, no DC drag field is applied to assist in longitudinal ion transport;
only residual gas flow along the RF-funnel axis transports ions into the downstream chamber while
an applied RF-field creates a radially confining potential.
This method was originally designed to extract ions from an He-buffer gas;
here, ions of mass 136\,u are to be extracted from Xe gas of very similar or equal mass.
The extraction of ions with mass equal to that of the carrier gas has never been tried for m\,$>40$\,u.
Detailed gas-dynamic and ion-trajectory Monte-Carlo simulations have been performed to optimize the design of the RF-only ion-funnel device for this purpose.
This design was initially reported in \cite{Bru13}.
Some results of these simulations are discussed in Section\,\ref{sec:simulations}.

\begin{figure}[t]
	\centering
		\includegraphics[width=\columnwidth]{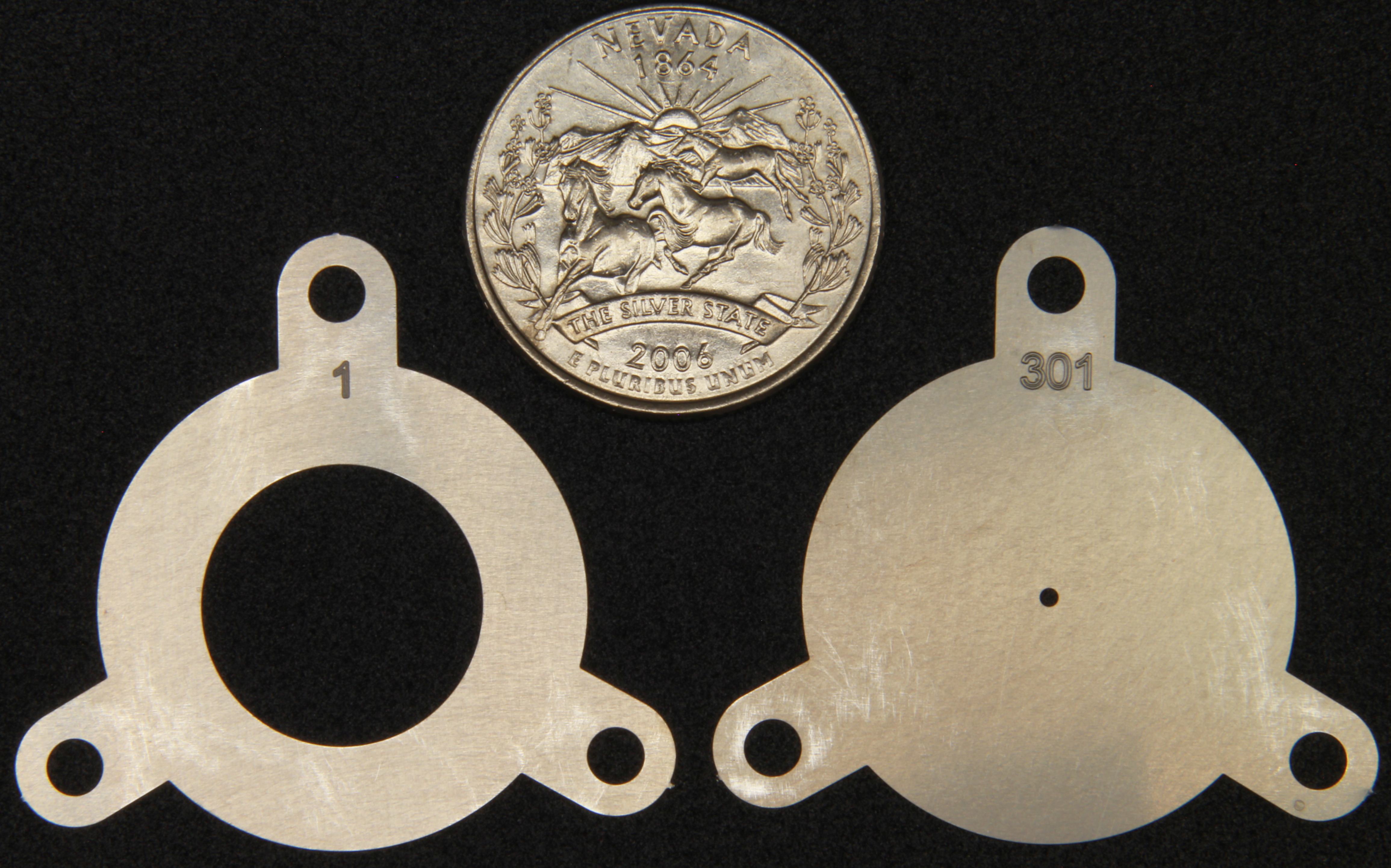}
	\caption{(color online)
Picture of the first (ID 16.0\,mm) and last electrodes (ID 1\,mm).
A\,US\,quarter\,(OD\,24.3\,mm)\,is shown\,for\,scale.%
}
	\label{fig:electrode}
\end{figure}

The RF-funnel consists of 301 individual electrode foils made from 0.102(3)\,mm thick high-tolerance, stainless steel sheets\footnote{\label{foot:sheetmetal}316 stainless steel alloy, ESPI Metals}.
The electrodes and inter-electrode spacers were manufactured using photo-etching\footnote{Newcut Inc. New York} to obtain high tolerances and low stress.
The electrodes are annular with outer diameter (OD) of 28\,mm and an inner diameter (ID) decreasing from 16.0\,mm to 1.0\,mm in constant steps of 0.05\,mm per electrode.
The electrodes have three mounting tabs that provide maximal spatial rigidity in the stack and allow for the required tolerances.
Prior to installation, each electrode was checked for flatness against a polished stainless steel surface and was flattened as required.
Each electrode is individually numbered for ease of assembly, as shown in Figure\,\ref{fig:electrode}.

\begin{figure}[t]
	\centering
		\includegraphics[width=\columnwidth]{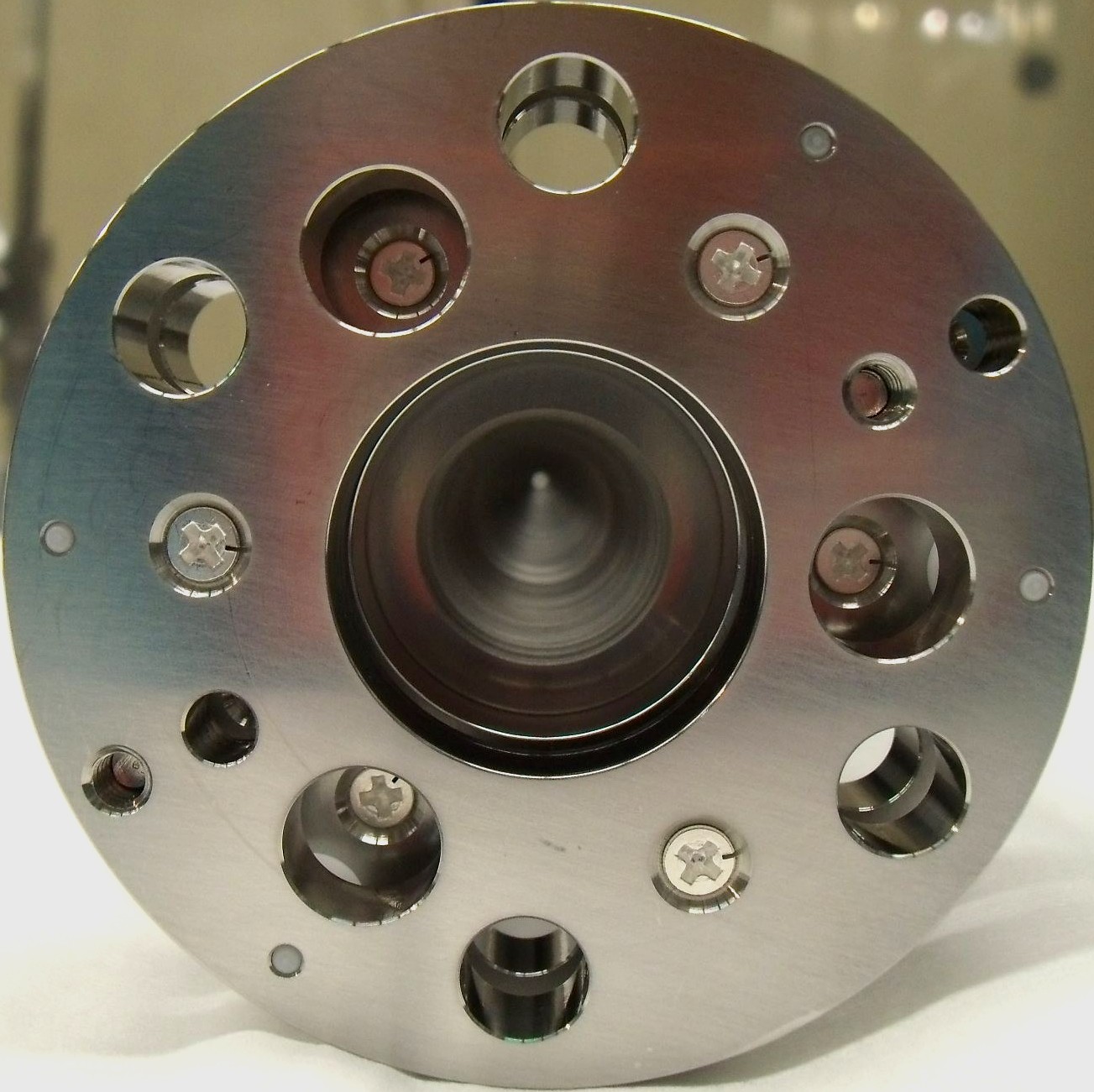}		
	\caption{(color online)
Picture of RF-funnel assembly prior to installation on the nozzle, view from base to exit.
The conical cavity ending with diameter 1\,mm (ID of electrode \#301) is visible.
Figure\,\ref{fig:funnel-assembly} shows this assembly (left) mounted to the nozzle (right).%
}
	\label{fig:picture-funnel-cone}
\end{figure}

Alternating electrodes were fed onto each of two electrically isolated sets (odd and even numbers) of mounting rods, 60$^{\circ}$ apart from each other.
Each set consists of three, 3\,mm diameter rods spaced equidistantly 16\,mm off the funnel axis.
On each set (odd or even) the electrodes are spaced by 0.610${\scriptsize\begin{array}{l}+0.005\\ -0.010\end{array}}$\,mm thick stainless steel spacers\textsuperscript{\ref{foot:sheetmetal}} of OD 5\,mm that were manufactured using photo-etching to achieve the required tolerances.
This results in an average separation of 0.25\,mm between the faces of neighboring electrodes (even to odd).
The conical cavity formed by decreasing electrode ID and the slot-vented screws that mount the supports are visible in Figure\,\ref{fig:picture-funnel-cone}.
After all electrodes were mounted, a glass-ceramic spacer was added after the last electrode to fix the radial position of the six mounting rods.
End caps are installed on each mounting rod to provide constant pressure to the electrode stacks.
\begin{figure}[t]
	\centering
		\includegraphics[width=\columnwidth , trim = 0 11 0 1, clip = true]{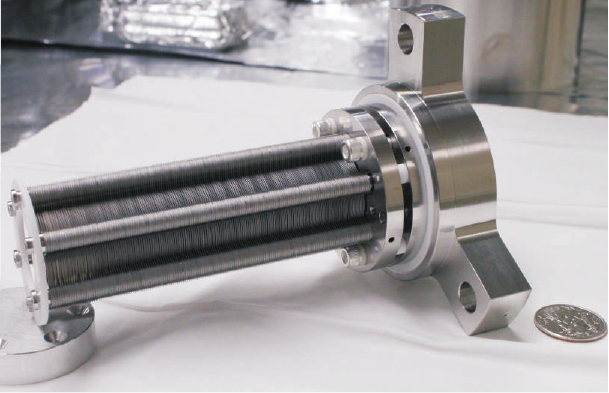}
	\caption{(color online)
Picture of assembled RF-funnel mounted on the nozzle flange.
Gas is injected from right.%
}
	\label{fig:funnel-assembly}
\end{figure}

The RF-funnel assembly is a stand-alone unit that is mounted onto and electrically isolated from the downstream side of the nozzle flange.
Figure\,\ref{fig:funnel-assembly} shows the nozzle-funnel assembly consisting of the nozzle-flange and RF-funnel assembly;
the electrically insulating glass-ceramic disk is visible at the RF-funnel's base (center-right).
The distance between the exit of the diverging nozzle and the first funnel electrode is 0.25\,mm.

\begin{figure}[t]
	\centering
	\includegraphics[width=\columnwidth]{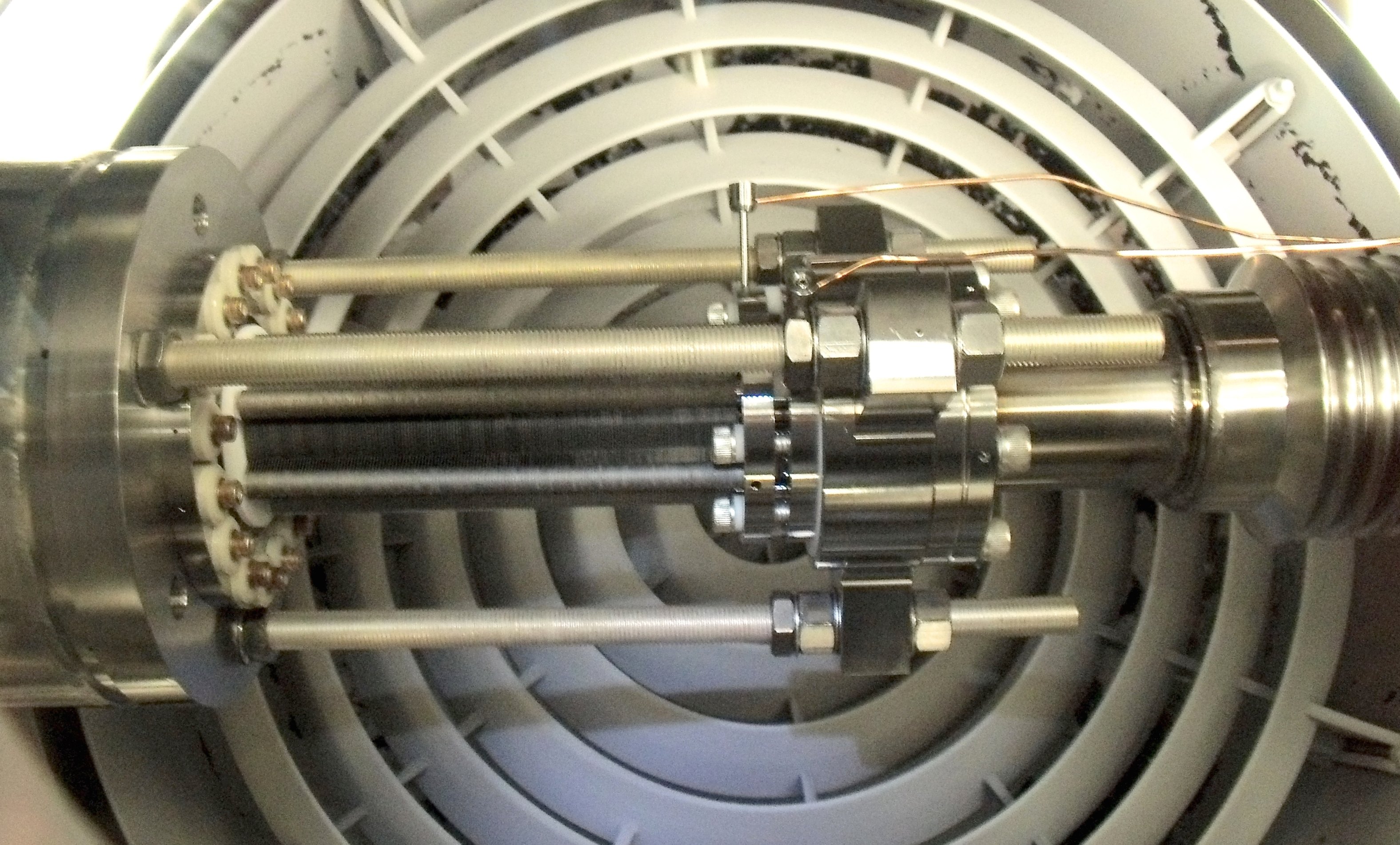}		
	\caption{(color online)
Picture of RF-funnel in chamber B above cryopump during a run.
Frozen xenon appears as a white coating.%
}
	\label{fig:picture-funnel}
\end{figure}

Figure\,\ref{fig:picture-funnel} shows a picture of the RF-funnel assembly installed inside chamber B between downstream chamber C (left) and high pressure chamber A (right).
The nozzle-funnel assembly is mounted onto chamber C by three silver-plated threaded rods (3/8$''$-24).
This assembly was aligned with respect to chamber C by adjusting the position of the nozzle flange mounting ears (center in Fig.\,\ref{fig:picture-funnel}) on the mounting rods at STP.
The last funnel electrode and aperture\,1 are coaxial (with under 0.03\,mm eccentricity) and separated by 0.233(8)\,mm.
Aperture\,1 is sealed to the surface of the downstream chamber C with an O-ring.
Chamber A is mounted on the upstream side of the nozzle-funnel assembly.
A bellows is welded around chamber A to compensate for tolerances in the assembly and mounts to chamber B via an inverted conflat flange.
A section view of an engineering model is shown in Figure\,2 of \cite{Bru13}.
Visible below the RF-funnel are metal louvers in the cryopump's first stage with a white coat of frozen xenon.

The capacitance of the RF-funnel installed in chamber B is found 
to be 6.014\,nF at $3.6$\,$\cdot$\,$10^{-7}$\,mbar when the cryopump is off.
This has a capacitive reactance of $\sim$10\,$\Omega$ at the typical operating frequency of 2.6\,MHz.
A frequency generator 
supplies a sinusoidal waveform to a broadband amplifier\footnote{ENI A150}.
A 1:4 balun\footnote{Balun Designs Model 1413t 1:4/3kW} is used to impedance match.
Due to a maximum amplifier output power of 120\,W, the highest RF-frequency where 100\,V$_{\textnormal{PP}}$ can be coupled to the funnel is 2.6\,MHz.
The applied voltages are lower than the break-down voltages in argon and xenon, i.e., they are lower than the Paschen minimum.

A set of baffles (not pictured in Fig.\,\ref{fig:picture-funnel}) has been installed underneath the funnel to reduce radiative cooling by the cryopump.
They are designed to block direct sight between the two elements while allowing a high gas flow conductance.
All ion-extraction measurements presented in this work have been carried out in this configuration;
only the pressure measurements presented in Figure\,\ref{fig:Ar-gas-run} were recorded without it.
\subsection{SPIG}\label{sec:SPIG} 
The gas flow through the exit of the RF-funnel generates pressure in chamber C, P$_{\textnormal{C}}$, which is too high to operate a CEM safely.
Thus, chamber C is used as an additional differential pumping stage and ions are transported through to chamber D where the CEM is located.
As the funnel is centered in chamber B for maximum cryopumping speed, chamber C is 0.5\,m long.
During gas operation, the pressure in this chamber is on the order of a \textmu bar; for xenon this corresponds to a mean-free path of about 40\,mm.
The pressure is highest immediately after the entrance to C where the gas flow regime changes from viscous to molecular flow.
Thus, the ions have to travel a length corresponding to at least 10 mean-free-path lengths to cross C, which requires an ion transport robust to collisions.
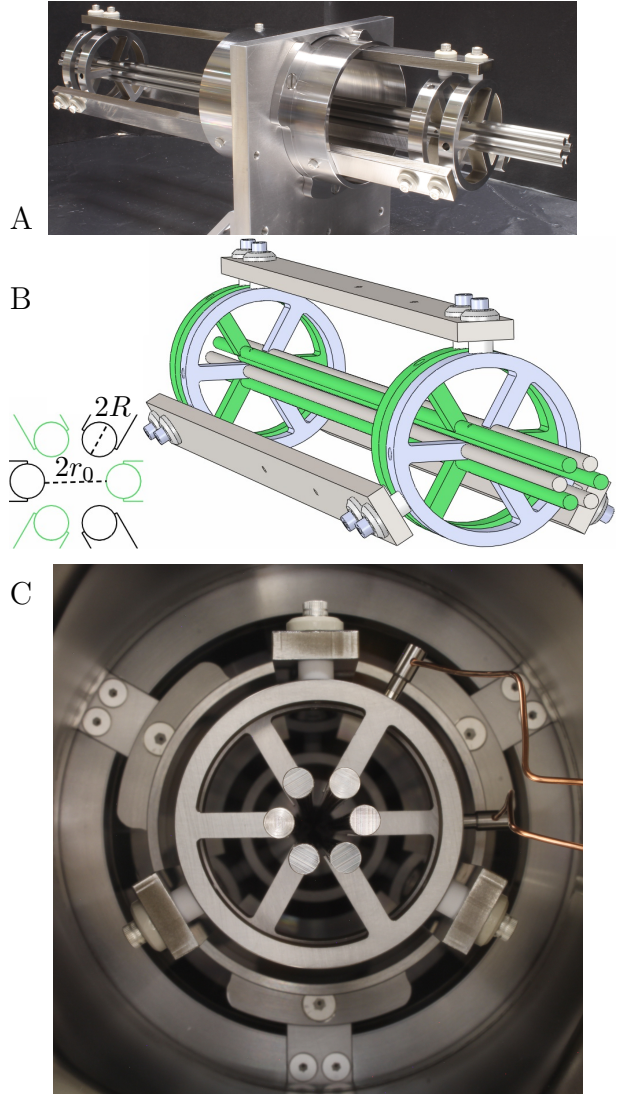
\begin{figure}
	\centering
		\hspace*{0pt}\includegraphics[trim = 0pt 420pt 0pt 0pt, clip, width=200pt]{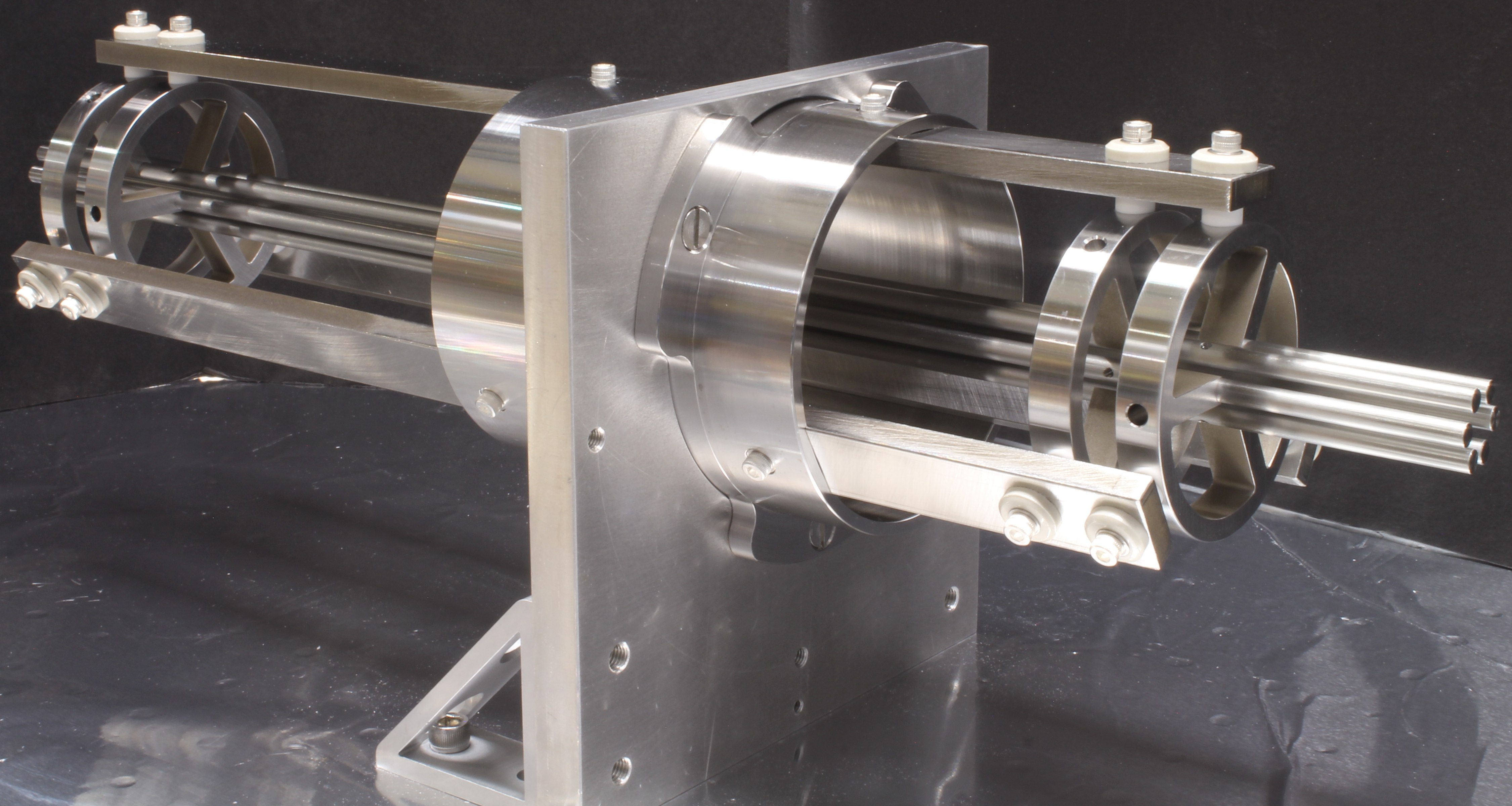} 
		\hspace*{0pt}\includegraphics[width=\columnwidth]{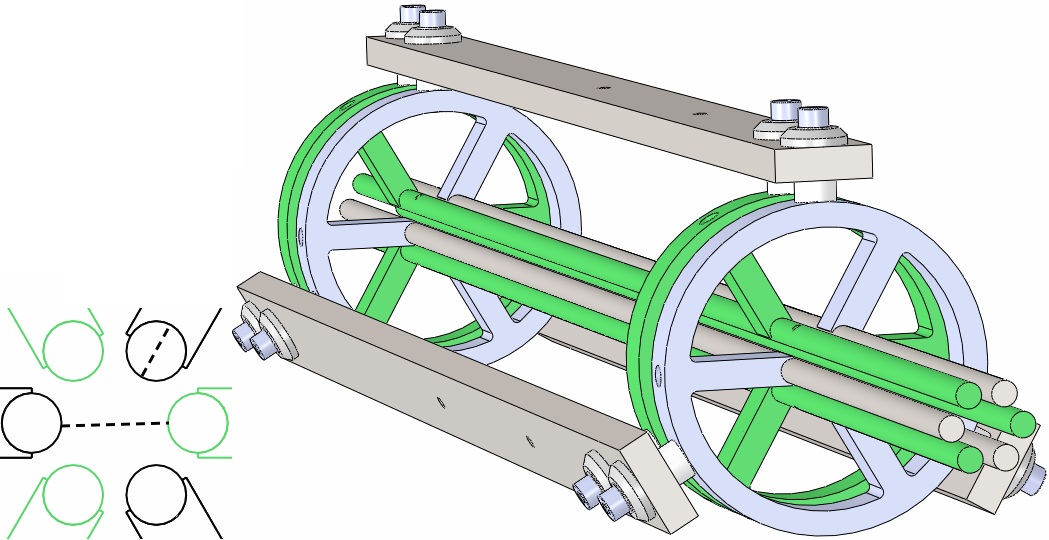}	
		\begin{picture}(229,0)
			\put(17,42){\small 2$r_0$}
			\put(32,65){\small 2$R$}
		\end{picture} 
		\vspace*{-24 pt}
		\begin{picture}(229,0)
			\put(0,150){A}
			\put(0,120){B}
			\put(0,8){C}
		\end{picture}
		\vspace*{-6pt} 
		\hspace*{0pt} \includegraphics[width = 200pt]{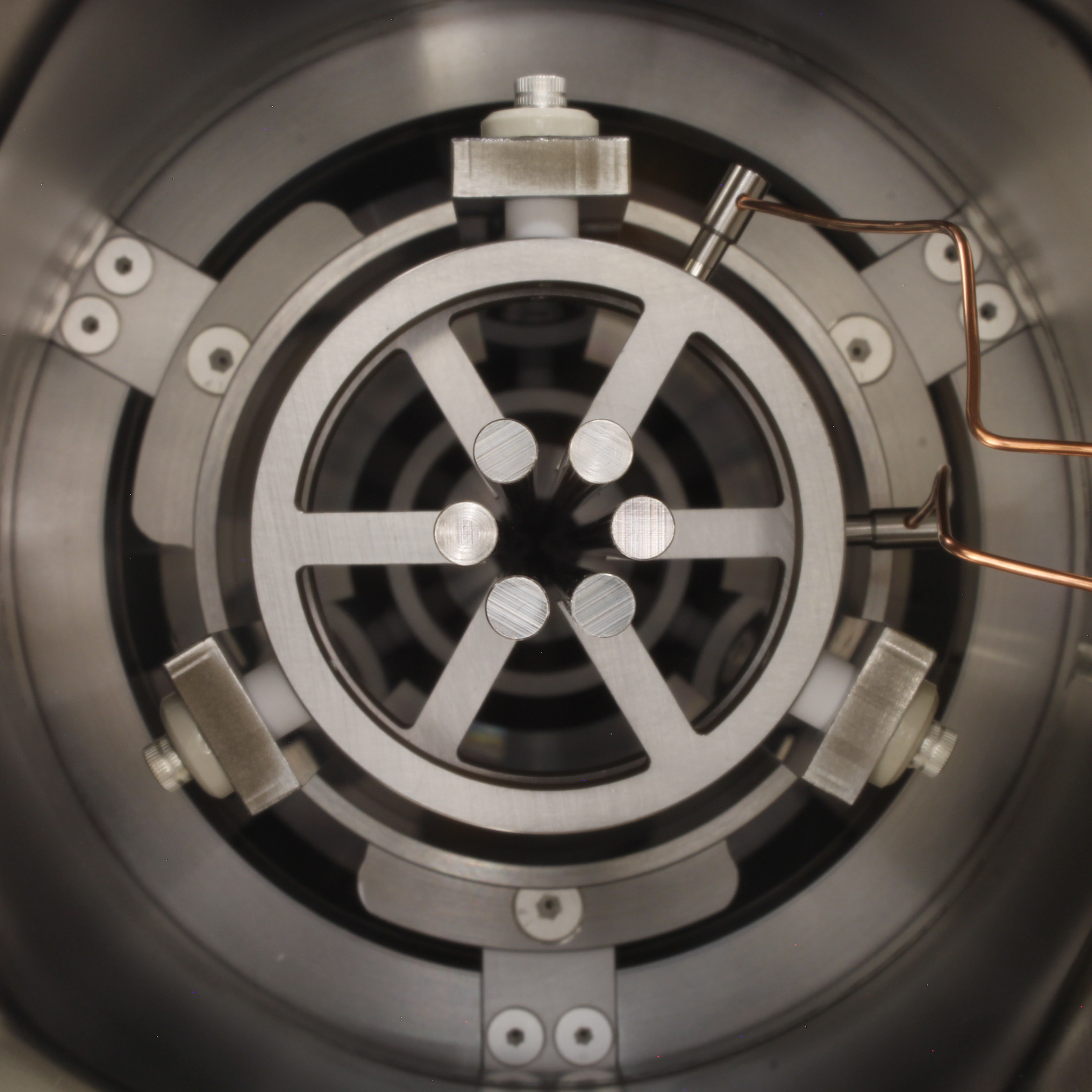} 
	\caption{(color online) 
(A) Picture of the initial SPIG assembly. 
(B) Rendering of the SPIG's two electrode sets (one set colored green).
Each set consists of three rods mounted between two circular positioning holders.
(C) View down the installed SPIG.%
}
	\label{fig:SPIG-pic}
\end{figure}

In RF-ion guides, losses of ions lighter than the buffer gas are expected to occur through RF-heating \cite{Maj68}.
However, for ions of mass equal to that of the background gas, e.g. Ba in Xe, on average no RF-heating is expected to occur \cite{Schwarz08}. 
For equal ion masses, losses will still occur from non-Gaussian fluctuations when consecutive collisions occur at RF-heating maxima \cite{DeVoe09}. 
This heating only happens near the electrodes where the electric field is larger \cite{Maj05}. 
The effective potential of a multipole guide, $\Phi_{\mbox{eff}} (r)\propto \left(r/r_0\right)^{N-2} ,$ 
has radial dependence of order two less than the number of poles, $N$,
thus guides with a higher number of poles exhibit a flatter potential at the center of the ion guide (for the same inscribed diameter $2r_0$) \cite{Gerlich1992}.
This minimizes the RF heating over most of the transport volume.
On the other hand, for a given $r_0$, the rod diameter $\left(2 R\right)$ decreases with increasing number of poles.
Thus the mechanics of mounting multipoles increase in complexity;
in particular, if the 0.5\,m long rods cannot support themselves.
These opposing considerations led to the choice of a sextupole ion guide (SPIG) for ion transport through chamber C.

Multipoles were simulated in SIMION \cite{Dahl00} with pressure response added from pre-built hard-sphere collisions model\footnote{\href{http://simion.com/info/collision_model_hs1.html}{\nolinkurl{simion.com/info/collision_model_hs1.html}}}. 
The most conservative estimate of pressure at the entrance of C, with the least favorable ion transmission efficiency, was used; 
this was obtained by solving the Navier-Stokes equations\footnote{SolidWorks 2012 Flow Simulation} in the entrance region.
In addition to the RF, a DC bias was determined to be necessary to collect the ions into the SPIG.
SIMION calculations showed that without a longitudinal DC potential gradient (i.e. with equal DC bias at  both ends of the SPIG), ion-transmission efficiencies of up to 0.8 can be achieved, and so the additional complexity of implementing a DC potential was avoided.

Great care was taken in the design of the SPIG to ensure mechanical rigidity and to create well defined reference features with tolerances less than 0.1\,mm. 
The SPIG (see Fig.\,\ref{fig:SPIG-pic}) consists of two electrode sets, each including three 480.5\,mm long rods of $2R = 4.76$\,mm (3/16$''$). 
These sets are mounted at both ends on circular positioning holders that maintain the rods 120$^{\circ}$ apart on an inscribed circle of radius $r_0 = 4.23$\,mm from the axis (see Fig.\,\ref{fig:SPIG-pic}).
This corresponds to the optimum ratio of the radius of the rods to the radius of the circle inscribed between the rods of 0.563 for the ideal sextupole field using cylindrical rods \cite{Konenkov2010}. 
The two sets of rods are mounted 60$^{\circ}$ apart onto three support bars.
The rod sets are insulated from the bars and positioned with precision ceramic spacers that insert into wells in the bar and holders.
The bars are mounted to a cylindrical center mount with three lobes that mount to a double-faced DN160 (CF$8''$) flange in chamber C to ensure that the SPIG is radially centered.
A picture of the SPIG mounted in chamber C is shown in Figure~\ref{fig:SPIG-pic}C, the lobes are visible at 60, 180, and 300 degrees.
The entrance to the SPIG lies 6.6\,mm after the aperture into chamber C (Aperture\,1 in Fig.\,\ref{fig:funnel-concept}); the aperture widens from 1\,mm diameter to 20.3\,mm over a distance of 5.58\,mm. 
The exit of the SPIG is 4\,mm in front of chamber D (Aperture\,2 in Fig.\,\ref{fig:funnel-concept}).

The installed SPIG has a capacitance between rod sets of 90\,pF at 10\,kHz. 
In order to drive the SPIG, a sinusoidal waveform is generated, 
amplified 
and coupled to the SPIG through a toroidal transformer.
The secondary of the transformer can be biased to float the SPIG to various DC voltages.
The SPIG is typically operated at 2.0\,MHz and 400\,V$_{\textnormal{PP}}$ with a -5.2\,V$_{\textnormal{DC}}$ bias.
\subsection{Ion detector}\label{sec:cem} 
A differential pumping barrier (Aperture\,2 in Fig.\,\ref{fig:funnel-concept}) separates chambers C and D downstream from the SPIG.
This aperture is electrically isolated by a ceramic disk and sealed by two O-rings.
This sandwich configuration allows biasing of the aperture to extract ions from the SPIG.
The aperture is tapered and its shape has been optimized using SIMION simulations for ion extraction into chamber D.
Downstream of this aperture a CEM\footnote{DeTech 2403} detects the ions.
Two lenses focus the extracted ions onto the CEM.
The front of the CEM is biased to $-2.4$\,kV, while the rest is at ground.
The CEM's signal is capacitively coupled to a high bandwidth amplifier\footnote{\label{foot:ortec-amps}ORTEC VT120A} and recorded\footnote{Stanford Research SR620\label{foot:counter}}.
\subsection{Data acquisition system}\label{sec:DAQ} 
A LabVIEW\footnote{National Instruments (NI), LabVIEW 2000} program controls and monitors the gas handling, vacuum systems and the cryopump; 
all data is recorded at 1\,Hz. 
Several pressure gauges\footnote{10\,kTorr (MKS 722B), 1\,kTorr (MKS 722B), \\\hspace*{17pt}100\,Torr (MKS 627D), 20\,mTorr (MKS 627D), \\\hspace*{17pt}and 100\,mTorr (MKS 690A read by MKS 270)} (Fig.\,\ref{fig:pumping-scheme}), CC1\footnote{\label{foot:CCG}MKS model 431 cold cathode gauge} and P1\footnote{\label{foot:pirani}MKS model 317 pirani gauge}\textsuperscript{,}\footnote{CC1 and P1 read by the same MKS 937A controller} are digitized\footnote{NI cFP-AI-112}. 
The controller\footnote{MKS 937B} that reads out CC3 and CC4\textsuperscript{\ref{foot:CCG}} as well as P2 and P3\textsuperscript{\ref{foot:pirani}} communicates via an RS485 connection. 
The proportional valve (PV) and the pneumatic block valves\footnote{AP Tech: AP3000SM} (BV1 and BV2) are controlled by NI modules\footnote{NI cFP-AO-210 and NI cFP-DO-400}.
A mass-flow meter\footnote{MKS 179A} (FM) measures xenon gas flow before the SAES purifier. 

A separate data acquisition program controls and reads the counter\textsuperscript{\ref{foot:counter}}, and simultaneously records pressure data. 
This program also controls the RF generator for the funnel, allowing scans of the applied RF amplitude.
\section{Ion extraction simulations} \label{sec:simulations}
\begin{figure*}
	\centering
		\includegraphics[width=0.91\textwidth]{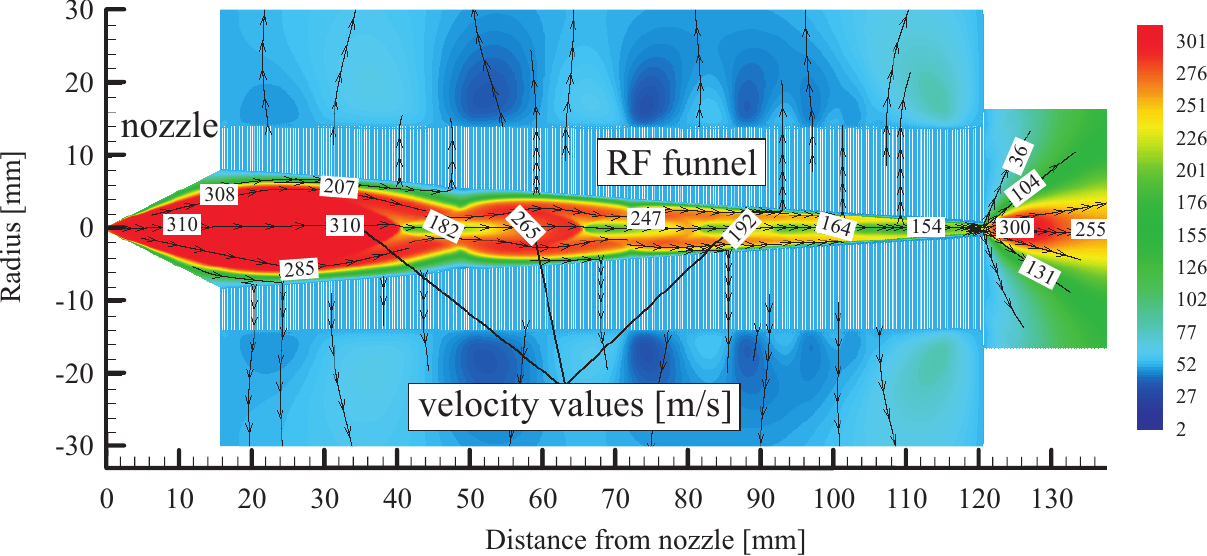}
	\caption{(color online) 
	Simulated velocity field for xenon with P$_{\textnormal{A}}=10$\,bar, P${_{\textnormal{B}}=8\cdot 10^{-3}}$\,mbar, and P$_{\textnormal{C}}=1.5\cdot 10^{-3}$\,mbar. The color scale indicates velocity in m/s.%
}
	\label{fig:Gas-Sim}
\end{figure*}
Numerical simulations, similar to \cite{Varentsov2004}, have been performed to optimize the design of the nozzle-funnel system for a xenon stagnation pressure in chamber A, P$_{\textnormal{A}}$, of 10\,bar with mass-flow restrictions posed by the cryopump.
These gas dynamic simulations resulted in nozzle and funnel dimensions presented in Section\,\ref{sec:funnel}.
The simulations are axisymmetric and ignore the electrode mounting features (legs, rods and washers), which might affect the conductivity of the funnel for vacuum pumping in chamber B.
Detailed information about gas flow fields of pressure, temperature, density, velocity, and Mach number in the nozzle and chambers A, B, and C has been obtained by means of the VARJET code \cite{Varentsov1998};
This code is based on the solution of a full system of time-dependent Navier-Stokes equations.
For the boundary condition in chamber B, a fixed background pressure, P$_{\textnormal{B}}$, is used.
This boundary was set at a radial distance of 50\,mm from the funnel axis, after comparison to a distance of 30\,mm showed no noticeable differences in the gas flow inside the funnel nor in ion transmission.

Gas dynamic calculations were performed for xenon gas (136\,u) and argon gas (40\,u).
The mass-flow rate through the nozzle equals 45.3\,mbar l/s at P$_{\textnormal{A}}=10$\,bar xenon and  at P$_{\textnormal{A}} = 5.4$\,bar argon.
The Reynolds number at the nozzle throat (the critical cross section) equals $6.4\cdot 10^{4}$ for P$_{\textnormal{A}} = 10$\,bar xenon and P$_{\textnormal{A}} = 18$\,bar argon;
here, the calculated gas flow shock-wave structures look the most similar.
The calculated velocity field for P${_{\textnormal{A}}=10}$\,bar xenon is shown in Figure\,\ref{fig:Gas-Sim}.

\begin{figure}
    \centering
    \includegraphics[width=\columnwidth]{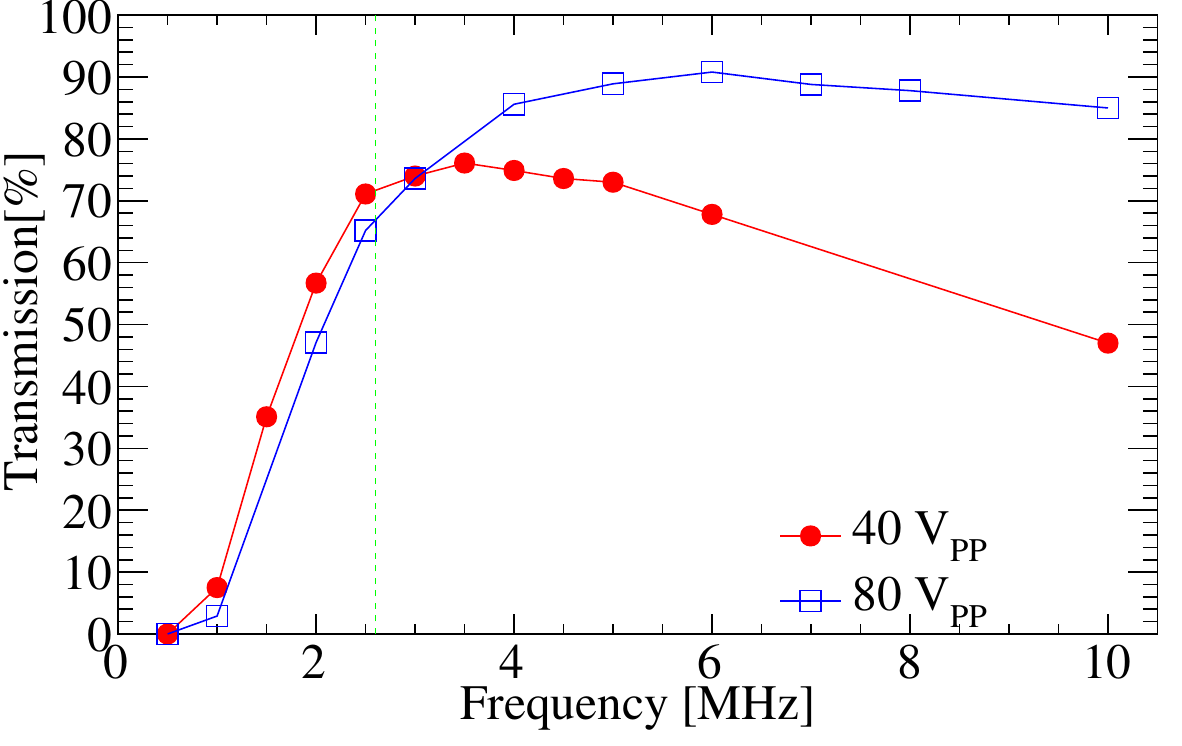}
	\caption{(color online)
Simulated, singly-charged, 136\,u ion transmission in xenon versus RF-frequency for 40 and 80\,V$_{\textnormal{PP}}$.
P${_{\textnormal{A}}=10}$\,bar, P$_{\textnormal{B}}= 3.5\cdot 10^{-3}$\,mbar, P$_{\textnormal{C}} =1.5\cdot 10^{-3}$\,mbar, nozzle diameter of 0.3\,mm and inter-electrode spacing of 0.3\,mm.
The vertical dashed line is at 2.6\,MHz, the frequency most commonly used.%
}
	\label{fig:transmission-amplitude}
\end{figure}

The results of these gas dynamic calculations were used as input to ion-trajectory Monte-Carlo simulations.
The ion transport efficiency of the RF-funnel was determined as the number of ions passing through the 1\,mm diameter exit electrode versus the total number of ions injected into the nozzle.
Typically, between 1100 and 7500 simulated $^{136}$Ba$^{+}$ ions were injected into the funnel.
Initial simulations with 10\,bar xenon were performed for selected frequencies between 0.5\,MHz and 10\,MHz.
The simulated transmission at 10\,bar xenon as a function of the applied RF-frequency is shown in Figure\,\ref{fig:transmission-amplitude}.

For increasing RF amplitudes, the maximum calculated transmission efficiency increases and occurs at an increased RF frequency.
However, most simulations were performed at a frequency of 2.6\,MHz which is typically used in measurements.
The use of 136\,u for ion and gas is motivated by barium tagging; using instead natural xenon and barium for transport gas and ion, respectively, results in similar, but negligibly enhanced, transmission efficiencies.

\begin{figure}
    \centering
    \includegraphics[width=\columnwidth]{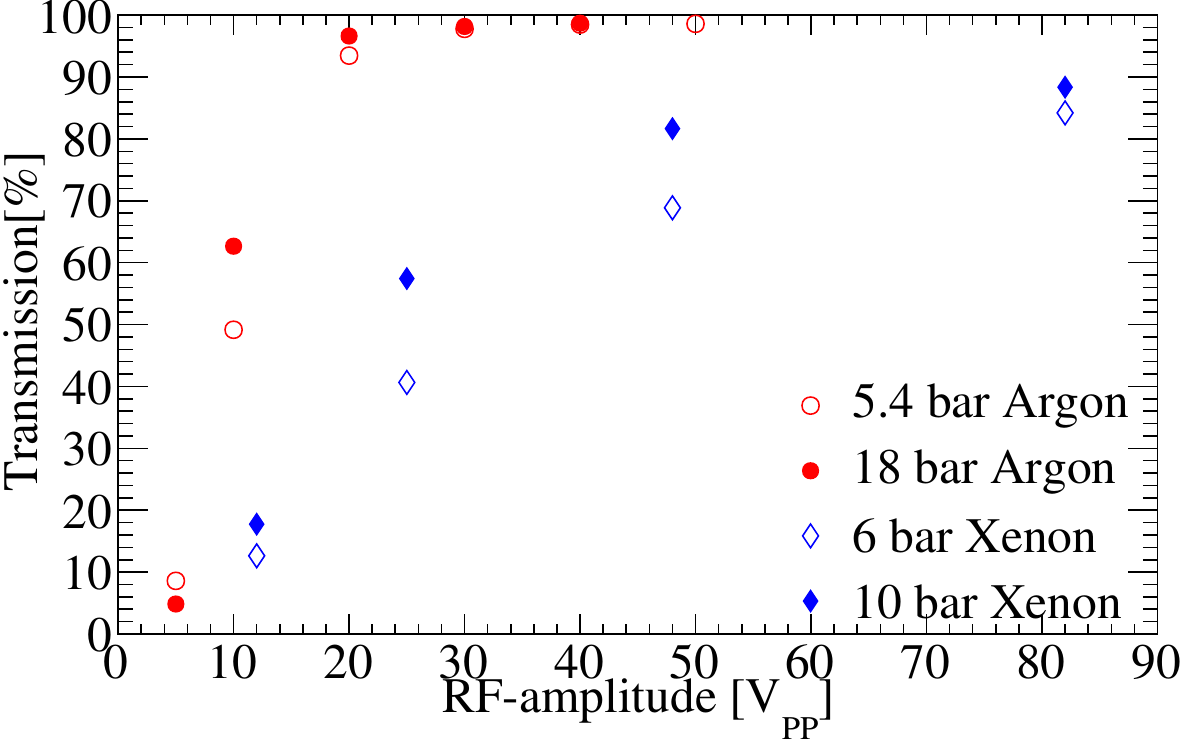}
	\caption{(color online)
Simulated transmission at P$_{\textnormal{B}}=3.5\cdot 10^{-3}$\,mbar and P${_{\textnormal{C}}=1.5\cdot 10^{-3}}$\,mbar, for P$_{\textnormal{A}}$ of 5.4\,bar and 18\,bar Ar and of 6\,bar and 10\,bar Xe with a nozzle diameter of 0.28\,mm.%
}
	\label{fig:Gas-Comp}
\end{figure}

The influence of RF-amplitude at 2.6\,MHz on transmission efficiency has been calculated for argon and xenon at selected P$_{\textnormal{A}}$ at P$_{\textnormal{B}}=3.5\cdot 10^{-3}$\,mbar and P$_{\textnormal{C}}=1.5\cdot 10^{-3}$\,mbar, as shown in Figure\,\ref{fig:Gas-Comp}.
With an argon transport gas, the large mass difference between $^{136}$Ba and Ar results in almost loss-free ion transport at RF amplitudes larger than $\sim$20\,V$_{\textnormal{PP}}$.
In the case of the equal-mass transport gas xenon, the losses inside the funnel increase.
Thus, higher RF amplitudes are required to create a stronger confining potential.
Furthermore, an increase in P$_{\textnormal{A}}$ at fixed P$_{\textnormal{B}}$ and P$_{\textnormal{C}}$ increases the gas flow and thus the transmission of $^{136}$Ba ions in xenon.
Comparisons of simulated RF-funnel behavior with data are presented in detail in the next section.
\section{Measurements}  
The measurements presented here focus on benchmarking the Monte-Carlo and gas dynamics calculations as well as on developing practicable modes of operation.
The key measurements performed include those of gas pressures during gas-jet operation, and ion extraction with respect to RF-funnel operation.
Measurements were performed using argon (Section\,\ref{Ar-studies}) or xenon (Section\,\ref{Xe-studies}) gas.

The pressure gauges used for the measurements (see Section\,\ref{sec:DAQ} and Fig.\,\ref{fig:pumping-scheme}) were found to agree within 10\,\% in common ranges.
For ion-extraction and transmission measurements the ion-count rate at the CEM (see Section\,\ref{sec:cem}) 
was read out with 1\,s integration time and then averaged over three to eleven values.
Only statistical uncertainties are shown for all plots. 
The CEM trigger threshold was optimized for each gas individually; this however, and the different ion production rates (see Figure\,\ref{fig:ion-production}) complicate objective comparison of measurements between argon and xenon gas operation.

Before each set of measurements, it was verified that all ions measured by the CEM originated in the ion source.
Firstly, the SPIG's RF amplitude was reduced to a few mV.
This eliminated the CEM signal, ruling out ion creation downstream of the SPIG.  
Secondly, the funnel's RF-amplitude was varied, 
altering the CEM-count rate, 
ruling out the production of secondary ions in the SPIG or chambers C or D. 
No indication of SPIG or RF-funnel sparking or arc discharge was observed.
\subsection{Systematic studies with argon\label{Ar-studies}}
\begin{figure}
	\centering
		\includegraphics[width=0.95\columnwidth]{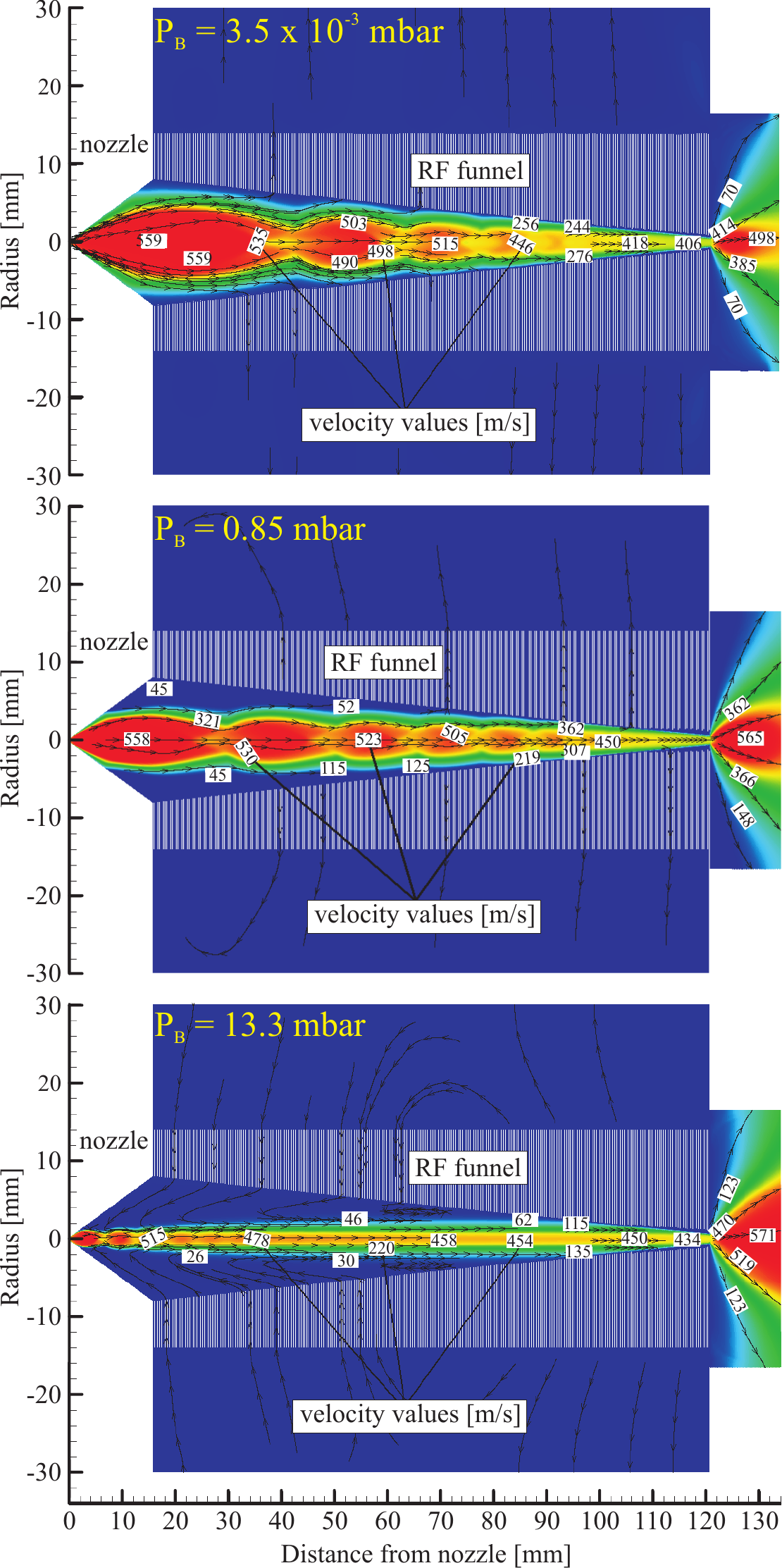}
	\caption{(color online)  
Simulated velocity fields with $P_{\textnormal{A}}=5.4$\,bar argon for background pressures P$_{\textnormal{B}}=3.5\cdot 10^{-3}$\,mbar (top), 0.85\,mbar (middle), and 13.3\,mbar (bottom).%
}
	\label{fig:Gas-Sim-PB}
\end{figure}

Gas-dynamic calculations show that the behavior of the flow inside the funnel depends on P$_{\textnormal{A}}$ as well as on the pressure surrounding the funnel, P$_{\textnormal{B}}$.
This effect has been investigated in detail at selections of P$_{\textnormal{A}}$ for P$_{\textnormal{B}}$ between $3.5\cdot 10^{-3}$\,mbar and 13.3\,mbar.
In calculations, P$_{\textnormal{B}}$ sets a boundary condition at 50\,mm off-axis in chamber B.
The velocity fields for an illustrative set of calculations for P$_{\textnormal{A}} = 5.4$\,bar argon at P$_{\textnormal{B}}=3.5\cdot 10^{-3}$\,mbar, 0.85\,mbar, and 13.3\,mbar is shown in Figure\,\ref{fig:Gas-Sim-PB}.
An increase in P$_{\textnormal{B}}$ results in the gas jet being closer to the axis of the funnel; this compression of the gas jet changes the gas flow and thus the ion transmission into chamber C.

\begin{figure*}
    	\centering
		\parbox{0.92\columnwidth}{\centering Calculated}\parbox{0.92\columnwidth}{\centering Measured}\\	
		\includegraphics[width=\textwidth]{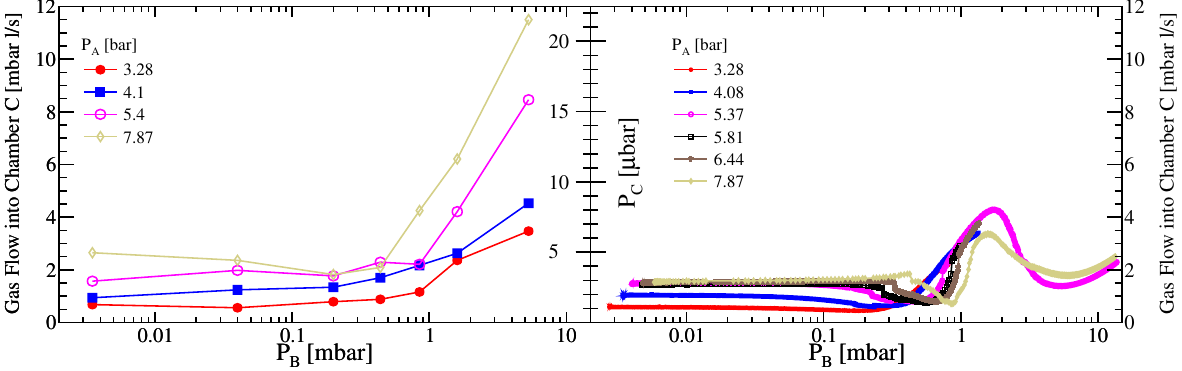}
	\caption{(color online) 
Calculated argon gas flow into chamber C and measured pressure in C as a function of background pressure, P$_{\textnormal{B}}$.
Selected argon stagnation pressures, P$_{\textnormal{A}}$, from 3.28\,bar to 7.87\,bar were applied.
A 534\,l/s Ar-pumping speed in C is used.%
}
	\label{fig:Gas-flow}
	\label{fig:Ar-gas-run}
\end{figure*}
Figure\,\ref{fig:Gas-flow} compares the calculated gas flow (left) into chamber C to the measured pressures (right).
The TMP on chamber C operates at a pumping speed of 534\,l/s at ${10\cdot 10^{-3}}$\,mbar argon \cite{EdwardsPump};
this nominal speed was used to convert between gas-flow and pressure.
The gas-flow rate into chamber C was calculated at a selection of P$_{\textnormal{A}}$ and P$_{\textnormal{B}}$.
For direct comparison, similar P$_{\textnormal{A}}$ were applied in the experiment.
The experimental measurement was performed dynamically, increasing P$_{\textnormal{B}}$ by virtue of the gradual heating of the cryopump subject to a long term ($\sim$30\,min) continuous gas load.
During each run, P$_{\textnormal{A}}$ was kept constant.
Runs and calculations were performed at P$_{\textnormal{A}}$ between 3.28\,bar and 7.90\,bar.%

For P$_{\textnormal{B}}\lesssim1.5$\,mbar the general trend in calculated and measured shape and pressure agrees within a factor of $\sim$2. 
However, the  calculated gas-flow rate for P${_{\textnormal{B}}\gtrsim 1.5}$\,mbar does not agree with the observed pressures; 
P$_{\textnormal{C}}$ steeply decreases with increasing P$_{\textnormal{B}}$ and has a second local minimum at $\sim$5\,mbar. 
This minimum is reproducible and present in all pressure measurements that were performed for background pressures up to 10\,mbar.
We conclude that in the regime (P${_{\textnormal{B}}\lesssim1.5}$\,mbar) where the funnel was optimized its  behavior is predicted by calculations.
At higher P$_{\textnormal{B}}$ thermal effects inside the vacuum chamber that are not included in calculations might cause the observed behavior.
In the following measurements, only the region P$_{\textnormal{B}}<1$\,mbar is considered.%
\begin{figure*}
	\centering
	\small
\parbox{\columnwidth}{\centering P$_{\textnormal{A}}=5.4$\,bar}\parbox{\columnwidth}{\centering P$_{\textnormal{A}}=7.87$\,bar}\\
\includegraphics[width=.49\textwidth]{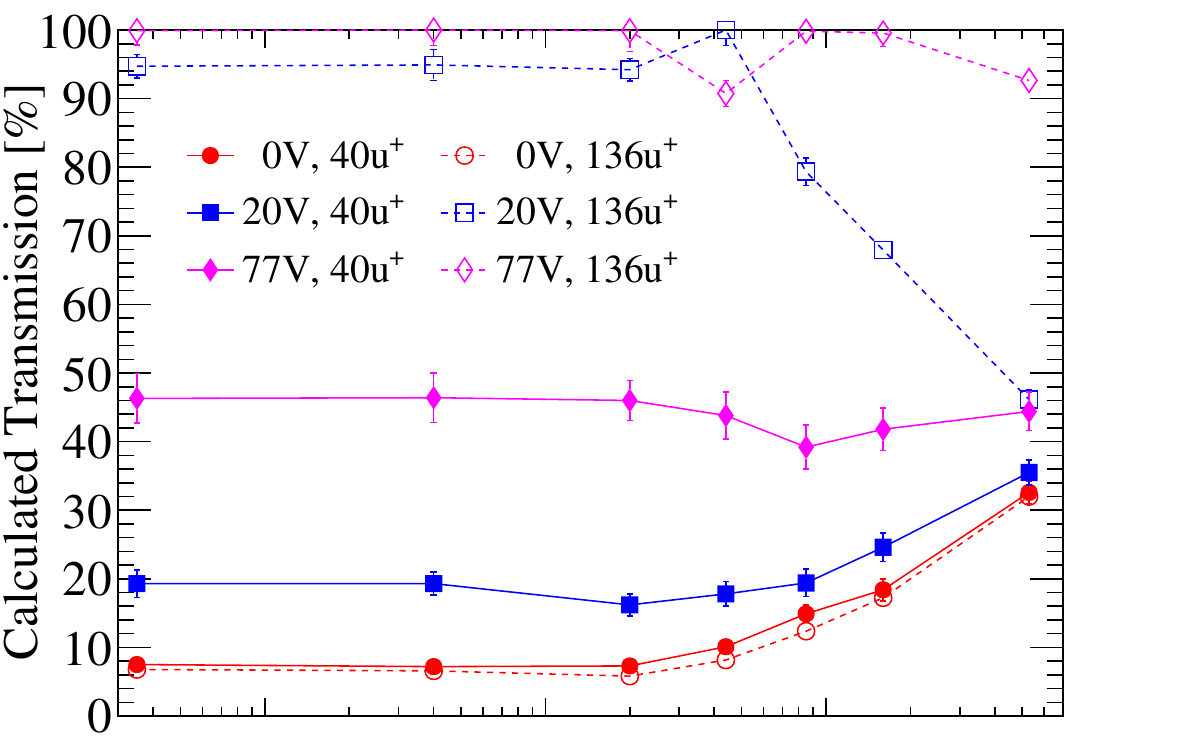}%
\includegraphics[width=.49\textwidth]{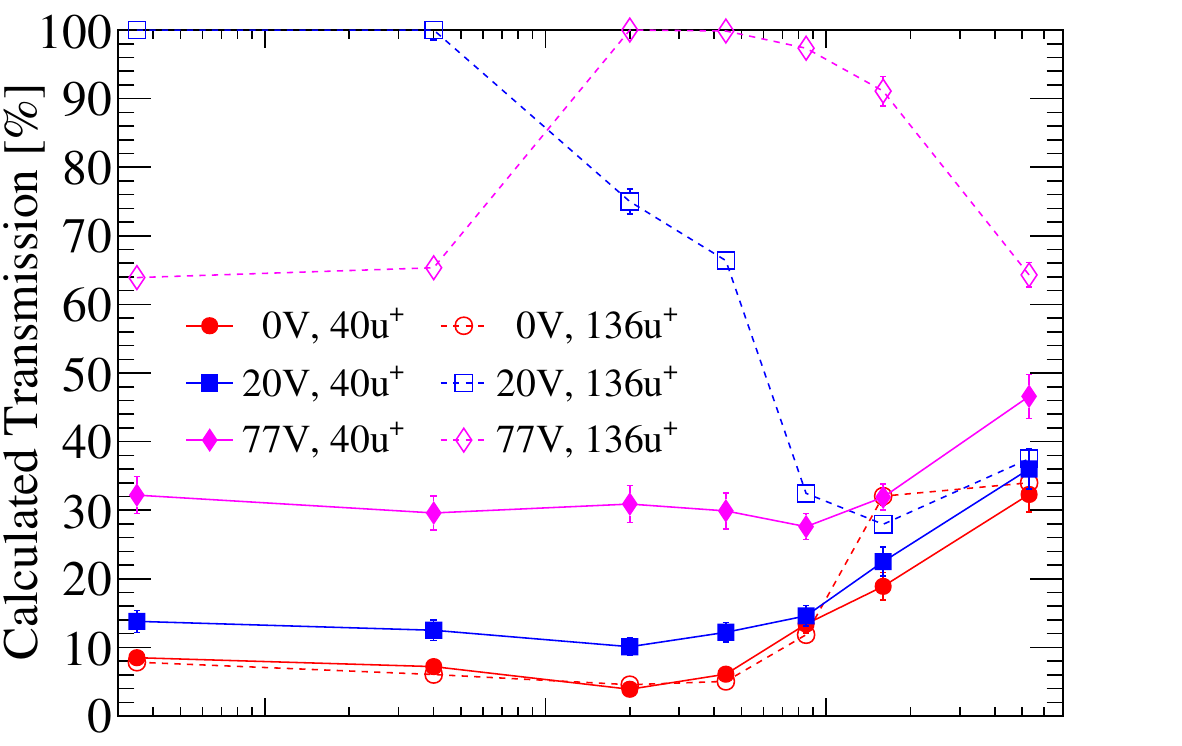}

\includegraphics[width=.49\textwidth]{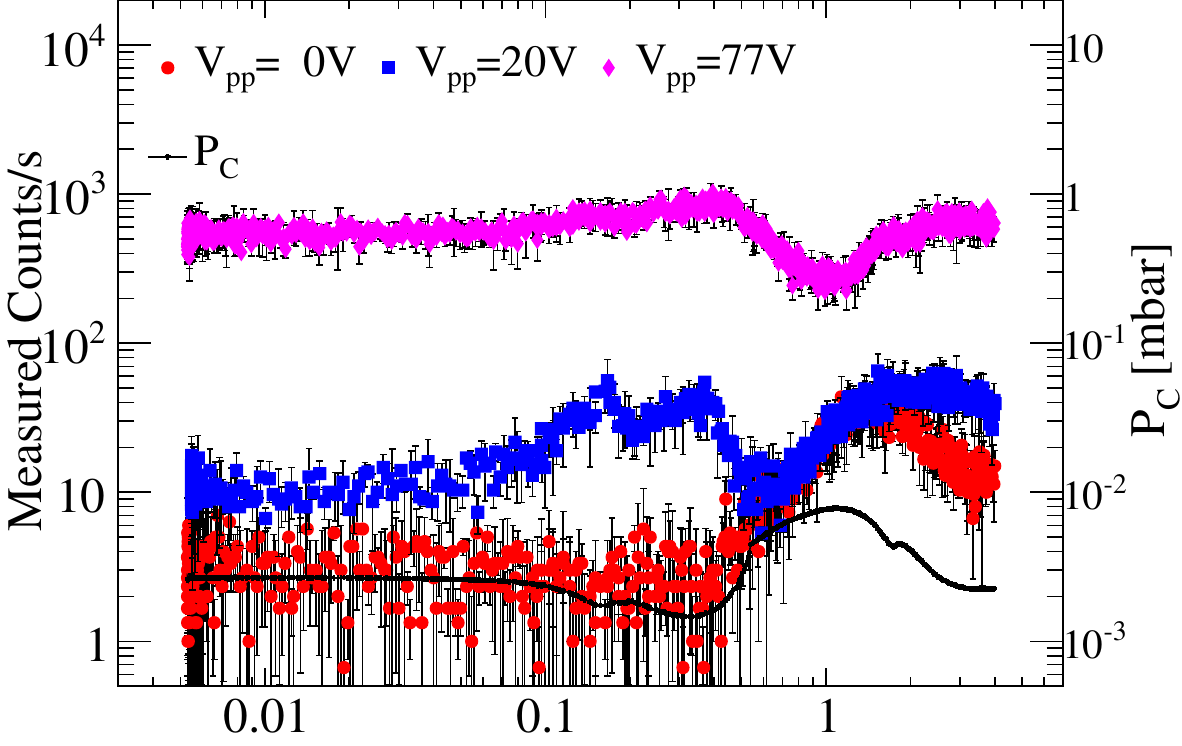}%
\includegraphics[width=.49\textwidth]{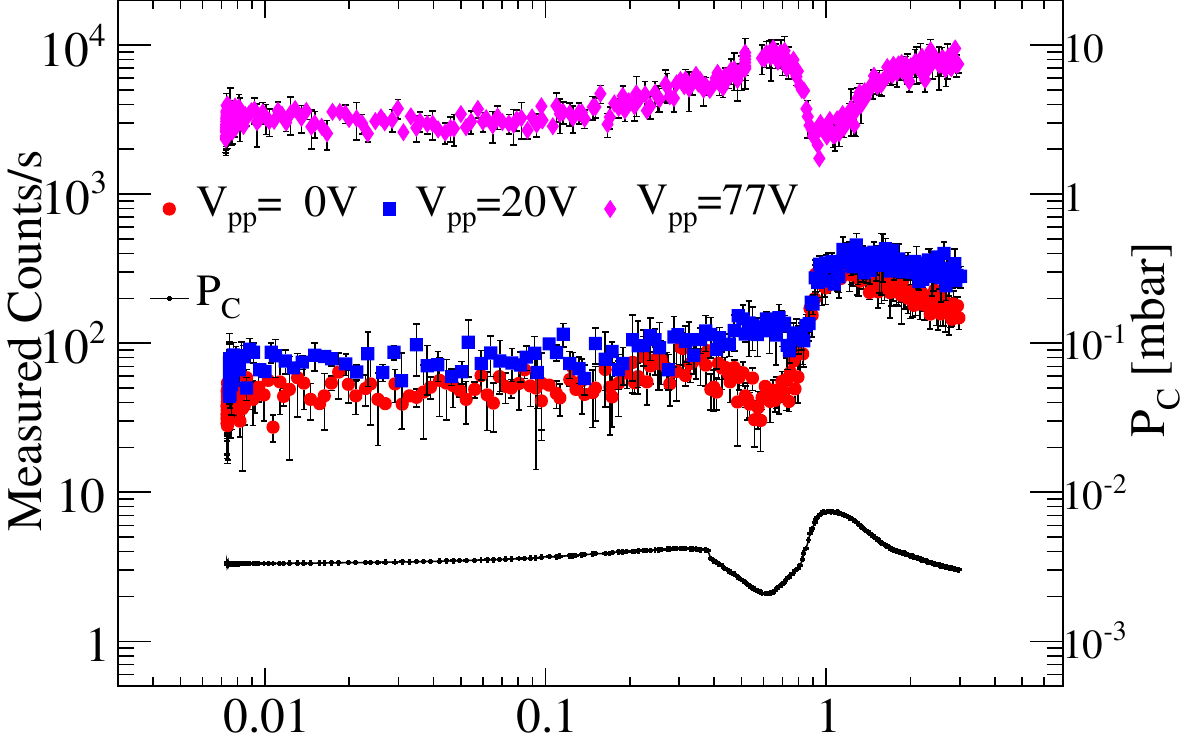}
\parbox{\columnwidth}{\centering P$_{\textnormal{B}}$ [mbar]}\parbox{\columnwidth}{\centering P$_{\textnormal{B}}$ [mbar]}
	\caption{(color online)
Calculated transmission (top) and measured count rate (below) at 5.36\,bar (left) and 7.87\,bar (right) argon stagnation pressure as a function of background pressure in chamber B. 
The RF-amplitude at 2.6\,MHz was set to 0\,V$_{\text{PP}}$ (red dot), 20\,V$_{\text{PP}}$ (blue square) and 77\,V$_{\text{PP}}$ (magenta diamond). 
The shown pressures, measured in chamber C during the 0Vpp run, are representative.
Calculations were performed with singly-charged ions of mass 40\,u and 136\,u.%
}
	\label{fig:Amplitude-comparison}
\end{figure*}

The ion-extraction efficiency in argon has been calculated for $^{40}$Ar$^+$ and $^{136}$Ba$^+$ and is shown in Figure\,\ref{fig:Amplitude-comparison}. 
The ion transmission was calculated for selected background pressures for P$_{\textnormal{A}}=5.40$\,bar and 7.87\,bar and RF amplitudes 0\,V$_{\textnormal{PP}}$, 20\,V$_{\textnormal{PP}}$, and 77\,V$_{\textnormal{PP}}$. 
Without any applied RF voltage (0\,V$_{\textnormal{PP}}$), ions are flushed into the downstream chamber C by the gas flow. 
Here, calculated ion transmissions are comparable for Ar$^+$ and Ba$^+$ and increase with the gas flow rate (see Fig.\,\ref{fig:Gas-flow}). 
Applying RF voltage to the funnel boosts transmission of Ba$^+$ to almost unity. 
The transmission of Ar$^+$ is reduced as expected due to the equal mass transport gas and the corresponding increased loss rate from collisions.
Higher RF-amplitudes generally led to better confinement of the ion and thus result in higher ion transmission. 

Measurements were performed applying the same parameters as those used in calculations of Figure\,\ref{fig:Amplitude-comparison}.
The measured ion-count rate along with pressures P$_{\textnormal{C}}$ and P$_{\textnormal{D}}$ for selected pressures P$_{\textnormal{B}}$ is shown at the bottom of Figure\,\ref{fig:Amplitude-comparison}. 
Due to the nature of the ion source, Ar$^+$ is expected to be the dominant ion species (see Section\,\ref{sec:ionsource}).

For an RF-amplitude of V$_{\textnormal{PP}}=0$\,V the ion count rate follows the gas flow rate and increases with increasing flow rate.
For low background pressures (0.01--0.1\,mbar), applying an RF amplitude of 20\,V$_{\textnormal{PP}}$ increases the  ion transmission by a factor of 3.5 (P${_{\textnormal{A}}=5.40}$\,bar) and 1.5 (P$_{\textnormal{A}}=7.87$\,bar) and is comparable to the calculated (0.04\,mbar, Ar$^{+}$) increase in transmission of 2.7 and 1.7.
In this pressure region, applying 77\,V$_{\textnormal{PP}}$ causes an increase in ion-count rate compared to 0\,V$_{\textnormal{PP}}$ of 170 and 78 for P$_{\textnormal{A}}=5.40$\,bar and P$_{\textnormal{A}}=7.87$\,bar, respectively;
this increase is higher than the calculated increase of 6.4 and 4.1.
\subsection{Systematic studies with Xenon\label{Xe-studies}}
\begin{figure}
	\centering
		\includegraphics[width=\columnwidth]{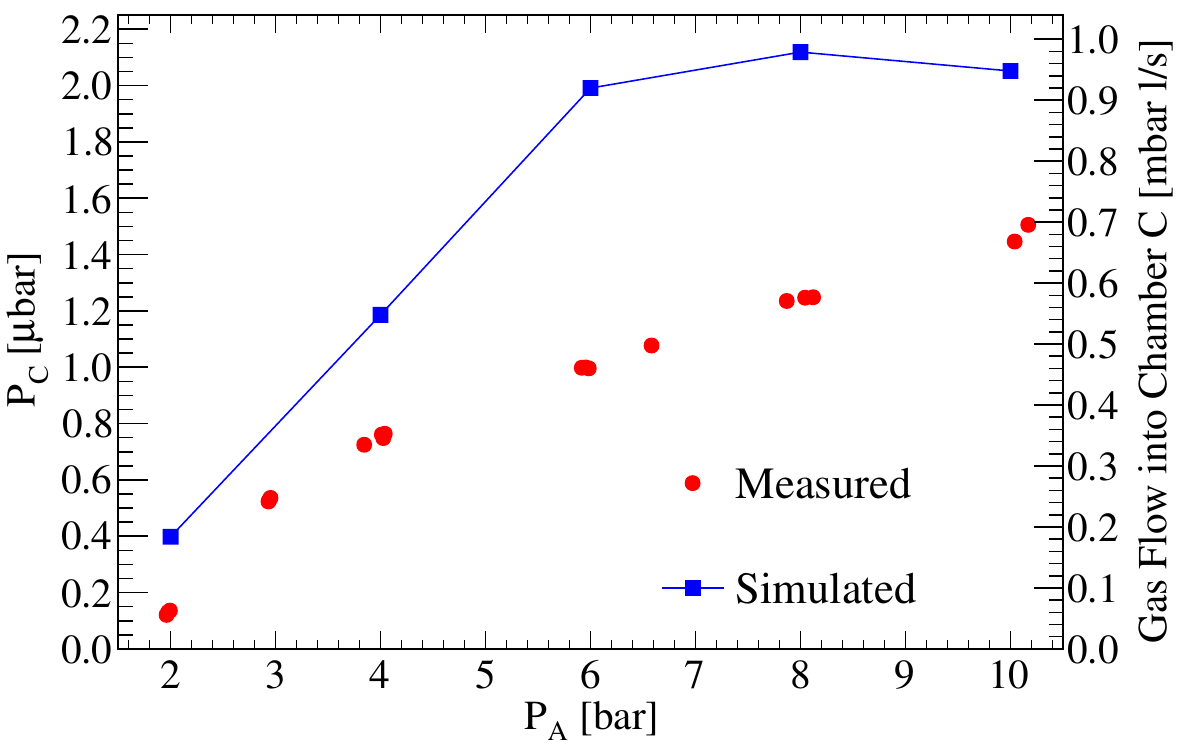}
	\caption{(color online) 
	Measured xenon pressure in chamber C (left axis) as a function of P$_{\textnormal{A}}$. 
	Calculated gas flow is shown on the right axis.
	The Xe-pumping speed is 462\,l/s.%
	}
	\label{fig:20140908-Xe-S-pressure-run-chamberC}
\end{figure}

The funnel operation was investigated for various P$_{\textnormal{A}}$ of xenon.
The calculated xenon-gas flow rates into chamber C for various stagnation pressures, using measured P$_{\textnormal{B}}$ for boundary conditions, are shown in Figure\,\ref{fig:20140908-Xe-S-pressure-run-chamberC} (right ordinate).
The measured pressures P$_{\textnormal{C}}$ are also shown in Figure\,\ref{fig:20140908-Xe-S-pressure-run-chamberC} (left ordinate).
The gas-flow rate into chamber C is obtained from P$_{\textnormal{C}}$ using a nominal xenon-pumping speed of 462\,l/s at $1.3\cdot 10^{-3}$\,mbar \cite{EdwardsPump}.
For increasing P$_{\textnormal{A}}$ the calculated flow rate increases before it levels off at $\sim$0.95\,mbar l/s for P$_{\textnormal{A}}\gtrsim 6$\,bar while the measured pressure continuously increases.
Despite this disagreement in the general shape of calculated gas flow into chamber C and the gas flow from measured pressure in chamber C, the calculated flow rates agree with the measured pressures to better than a factor of 2 under the assumed pumping speed.

\begin{figure*}
	\centering
		\parbox{\columnwidth}{\centering Calculated}\parbox{\columnwidth}{\centering Measured}\\
		\includegraphics[width=\columnwidth]{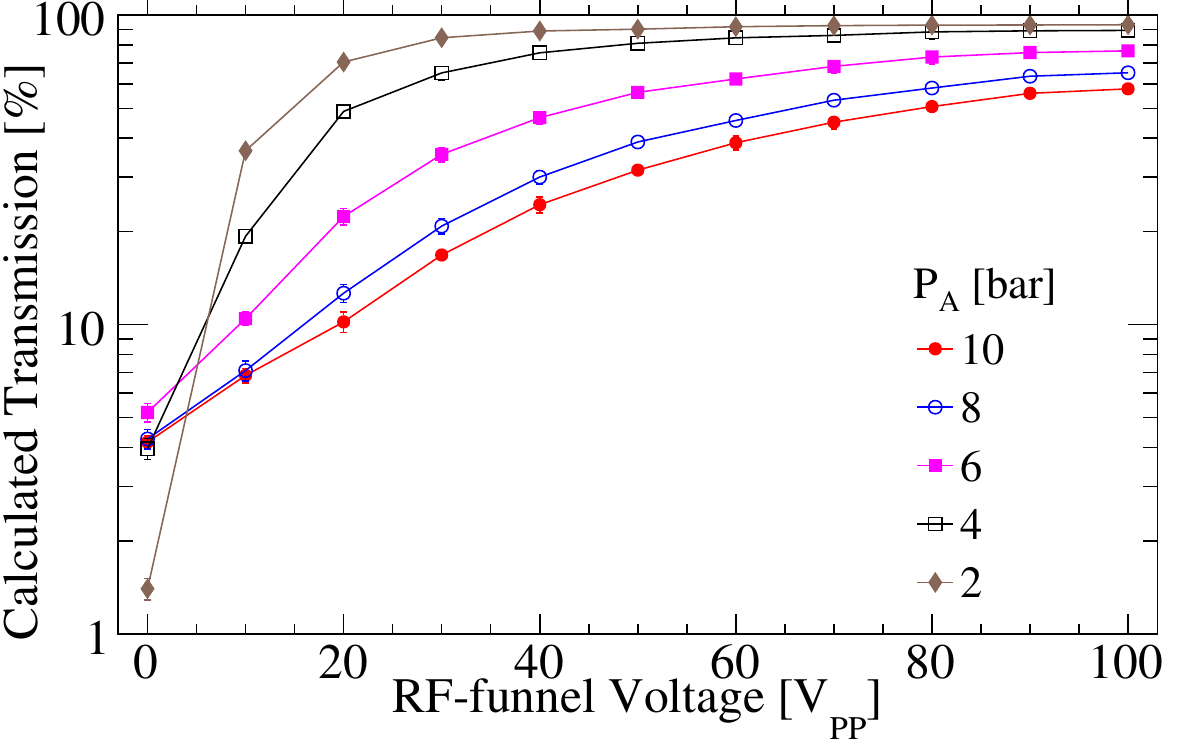}
		\includegraphics[width=\columnwidth]{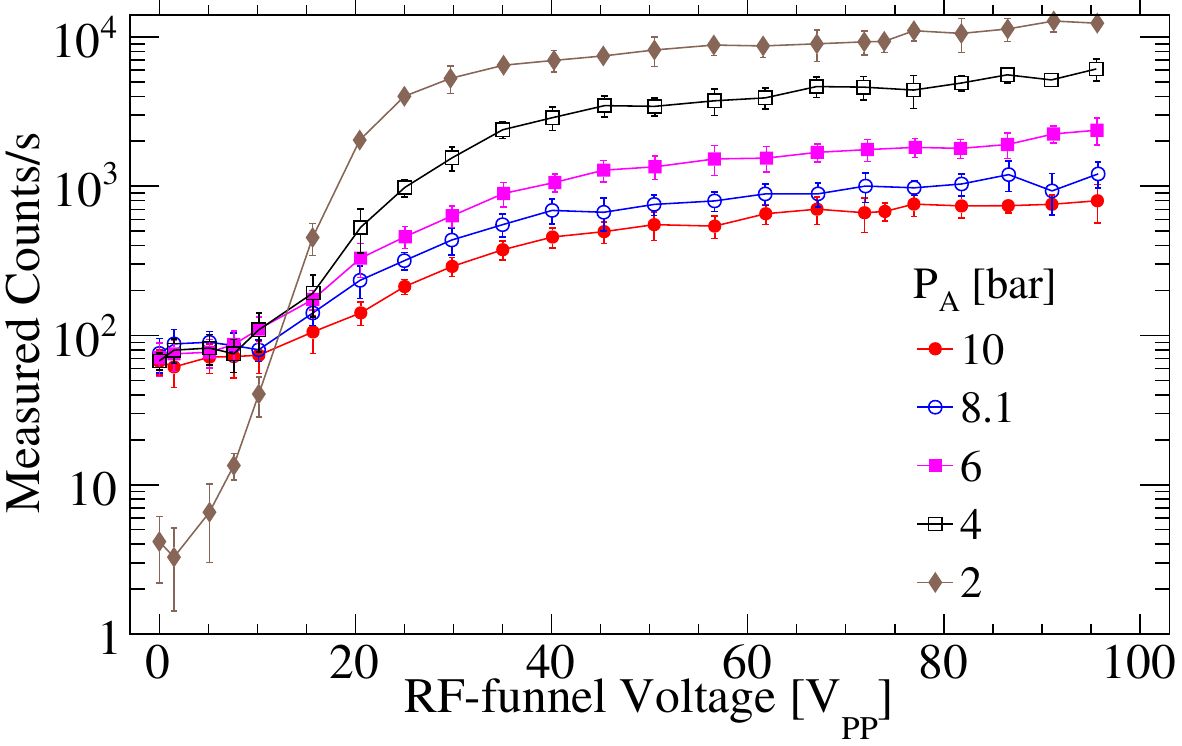}
	\caption{(color online)
	Calculated ion transmission (left) and observed ion-count rate (right) as function of RF-amplitude at 2.6\,MHz for selected xenon P$_{\textnormal{A}}$.
	In these calculations, measured P$_{\textnormal{B}}=8\cdot 10^{-3}$\,mbar and P$_{\textnormal{C}}=1.5\cdot 10^{-3}$\,mbar were used for boundary conditions.%
	}
	\label{fig:Xe-Amplitude-scan}
\end{figure*}

The effect of varying RF-amplitudes at 2.6\,MHz was simulated and measured for various P$_{\textnormal{A}}$ shown in Figure\,\ref{fig:Xe-Amplitude-scan}. 
The highest calculated ion transmission reaches 93\% at a xenon pressure of 2\,bar.
At this pressure, relatively low RF-amplitudes of $\gtrsim$30\,V$_{\textnormal{PP}}$ are sufficient to confine the Ba ions.
At higher P$_{\textnormal{A}}$, a given RF-amplitude is less efficient at confining ions.
At P$_{\textnormal{A}}=10$\,bar, the maximum transmission efficiency for 2.6\,MHz is calculated to be 58\%; 
higher transmission efficiencies have been calculated, reaching 85\% for 6\,MHz at 160\,V$_{\textnormal{PP}}$.
Due to hardware limitations, these RF settings could not be realized in this setup.

The measured ion-count rate in Figure\,\ref{fig:Xe-Amplitude-scan} for P$_{\textnormal{A}}=2$\,bar increases by up to 4.5 orders of magnitude while the calculated transmission only increases from 1.4(1)\% to 93.1(29)\%. 
However, the observed decrease in ion transmission at low RF-amplitudes is predicted by calculations.
Better agreement between calculated and measured ion-count rate is observed for higher P$_{\textnormal{A}}$.
Overall, the general trend and shape of both plots agree. 
It is pointed out, that count rates at different P$_{\textnormal{A}}$ cannot be compared quantitatively with each other due to the different ion production rates and the lack of ion identification.
Nevertheless, the shapes of calculated and measured ion extraction in Figure\,\ref{fig:Xe-Amplitude-scan} show similarities and indicate agreement.
\section{Conclusion and Outlook}
We designed and built a setup to extract ions from a high pressure noble gas. 
With this setup we demonstrated ion extraction from both xenon gas and argon gas at pressures up to 10\,bar. 
Experimental pressure data have been compared to gas dynamic calculations and agreement was found within a factor of 2.

An extension of the presented system is being developed to allow ion identification through a mass-to-charge ($m/q$) measurement technique, e.g. a quadrupole mass filter (QMF) or a 2-D Paul trap \cite{Schwartz2002}.
With such $m/q$ established, the extraction efficiency of Ba ions and other ions from high pressure noble gases will be investigated.
In a later step, the presented setup will be coupled to a buffer-gas filled Paul trap to unambiguously identify Ba ions through laser spectroscopy \cite{Green2007} and demonstrate the full process of ion extraction and spectroscopic detection.

Applications, other than Ba tagging are under investigation for RF-only funnels.
The combination of a laser with an RF-only funnel as an ion source has been proposed in \cite{Varentsov2008,Dieteker2012}.  
A selected species can be ionized using multiple lasers, which potentially allows contamination measurements in low-background noble gas detectors.
Such a system is best suited to measure contamination concentrations of isotopes heavier than the detector medium, e.g., radon in xenon gas or krypton in argon gas, and complements other contamination measurement techniques such as using an atom trap for trace analysis \cite{Chen1999,Jiang2011} or a cold trap \cite{Dobi2012}.

Another application for an RF-only funnel is the extraction of radioactive ions from stopper cells at fragmentation facilities.
Such a use is under investigation to improve ion extraction from the Cryogenic Stopping Cell for the Super-FRS at FAIR \cite{Ranjan2011}.

Since their introduction in the 1980s, RF-funnels have been improved constantly and find a wide application in mass spectrometry. 
Due to their high efficiency and simple application they are well suited for application in future high precision low background experiments.
\section{Acknowledgments} 
We thank
Maxime Brodeur (University of Notre Dame),
Stefan Schwarz (NSCL),
Peter Schury (RIKEN),
and
Mel Good (TRIUMF)
for fruitful discussions which were very helpful in the design and construction of the presented setup.
We acknowledge the professional help of the Stanford Machine shop, especially K. Merkle, and J. Kirk, as well as R. Conley for their help in machining parts.

This work was supported in the US by NSF grant PHY-0918469 and in the Russian Federation by grant RFBR 14-22-03028. 
\bibliography{references.bib}

\begin{thebibliography}{10}
\expandafter\ifx\csname url\endcsname\relax
  \def\url#1{\texttt{#1}}\fi
\expandafter\ifx\csname urlprefix\endcsname\relax\def\urlprefix{URL }\fi
\expandafter\ifx\csname href\endcsname\relax
  \def\href#1#2{#2} \def\path#1{#1}\fi

\bibitem{Kelly2010}
R.~T. Kelly, A.~V. Tolmachev, J.~S. Page, {et al.}, Mass Spectrom. Rev. 29
  (2010) 294.
\newblock \href {http://dx.doi.org/10.1002/mas.20232}
  {\path{doi:10.1002/mas.20232}}.

\bibitem{Reponen2011}
M.~Reponen, I.~Moore, I.~Pohjalainen, {et al.}, {Nucl. Instrum. Meth. A} 635
  (2011) 24.
\newblock \href {http://dx.doi.org/10.1016/j.nima.2011.01.125}
  {\path{doi:10.1016/j.nima.2011.01.125}}.

\bibitem{Page2006}
J.~S. Page, A.~V. Tolmachev, K.~Tang, R.~D. Smith, J. Am. Soc. Mass Spectr. 17
  (2006) 586.
\newblock \href {http://dx.doi.org/10.1016/j.jasms.2005.12.013}
  {\path{doi:10.1016/j.jasms.2005.12.013}}.

\bibitem{Page2007}
J.~S. Page, K.~Tang, R.~D. Smith, Int. J. Mass Spectrom. 265 (2007) 244.
\newblock \href {http://dx.doi.org/10.1016/j.ijms.2007.02.032}
  {\path{doi:10.1016/j.ijms.2007.02.032}}.

\bibitem{Ibrahim2006}
Y.~Ibrahim, K.~Tang, A.~V. Tolmachev, {et al.}, J. Am. Soc. Mass Spectr. 17
  (2006) 1299.
\newblock \href {http://dx.doi.org/10.1016/j.jasms.2006.06.005}
  {\path{doi:10.1016/j.jasms.2006.06.005}}.

\bibitem{Wada2013}
M.~Wada, Nucl. Instrum. Meth. B 317, Part B (2013) 450.
\newblock \href {http://dx.doi.org/10.1016/j.nimb.2013.08.062}
  {\path{doi:10.1016/j.nimb.2013.08.062}}.

\bibitem{Kim2001}
T.~Kim, K.~Tang, H.~R. Udseth, R.~D. Smith, Anal. Chem. 73 (2001) 4162.
\newblock \href {http://dx.doi.org/10.1021/ac010174e}
  {\path{doi:10.1021/ac010174e}}.

\bibitem{Anthony2014}
S.~N. Anthony, D.~L. Shinholt, M.~F. Jarrold, Int. J. Mass Spectrom. 371 (2014)
  1.
\newblock \href {http://dx.doi.org/10.1016/j.ijms.2014.06.007}
  {\path{doi:10.1016/j.ijms.2014.06.007}}.

\bibitem{Ibrahim2014}
Y.~M. Ibrahim, E.~S. Baker, W.~F. Danielson, III, et~al., Int. J. Mass
  Spectrom.~(0).
\newblock \href {http://dx.doi.org/10.1016/j.ijms.2014.07.034}
  {\path{doi:10.1016/j.ijms.2014.07.034}}.

\bibitem{Varentsov2001}
{V.L. Varentsov}, {A new approach to the extraction system design, SHIPTRAP,
  Collaboration Workshop, Mainz, Germany, March 19-20, 2001}, unpublished
  (2001).

\bibitem{Varentsov2002}
{V.L. Varentsov and D. Habs}, Nucl. Instr. and Meth. A 490 (2002) 16.

\bibitem{Varentsov2004}
{V.L. Varentsov and M. Wada}, Nucl. Instr. and Meth. A 532 (2004) 210.
\newblock \href {http://dx.doi.org/10.1016/j.nima.2004.06.078}
  {\path{doi:10.1016/j.nima.2004.06.078}}.

\bibitem{Avignone2008}
F.~T. Avignone, S.~R. Elliott, J.~Engel, Rev. Mod. Phys. 80 (2008) 481.
\newblock \href {http://dx.doi.org/10.1103/RevModPhys.80.481}
  {\path{doi:10.1103/RevModPhys.80.481}}.

\bibitem{Auger2012}
M.~Auger, D.~J. Auty, P.~S. Barbeau, et~al., Phys. Rev. Lett. 109 (2012)
  032505.
\newblock \href {http://dx.doi.org/10.1103/PhysRevLett.109.032505}
  {\path{doi:10.1103/PhysRevLett.109.032505}}.

\bibitem{Gando2013}
A.~Gando, Y.~Gando, H.~Hanakago, {et al.}, Phys. Rev. Lett. 110 (2013) 062502.
\newblock \href {http://dx.doi.org/10.1103/PhysRevLett.110.062502}
  {\path{doi:10.1103/PhysRevLett.110.062502}}.

\bibitem{Albert2014}
J.~B. Albert, D.~J. Auty, P.~S. Barbeau, {et al.}, Nature 510 (2014) 229.
\newblock \href {http://dx.doi.org/10.1038/nature13432}
  {\path{doi:10.1038/nature13432}}.

\bibitem{Agostini2013}
M.~Agostini, M.~Allardt, E.~Andreotti, {et al.}, Phys. Rev. Lett. 111 (2013)
  122503.
\newblock \href {http://dx.doi.org/10.1103/PhysRevLett.111.122503}
  {\path{doi:10.1103/PhysRevLett.111.122503}}.

\bibitem{Moe91}
M.~K. Moe, Phys. Rev. C 44 (1991) R931.
\newblock \href {http://dx.doi.org/10.1103/PhysRevC.44.R931}
  {\path{doi:10.1103/PhysRevC.44.R931}}.

\bibitem{Twelker2014}
K.~Twelker, S.~Kravitz, M.~{Montero D\'iez}, {et al.}, Rev. Sci. Instr. 85~(9).
\newblock \href {http://dx.doi.org/10.1063/1.4895646}
  {\path{doi:10.1063/1.4895646}}.

\bibitem{Mong2014}
B.~Mong, S.~Cook, T.~Walton, {et al.}, Phys. Rev. A91~(2) (2015) 022505.
\newblock \href {http://dx.doi.org/10.1103/PhysRevA.91.022505}
  {\path{doi:10.1103/PhysRevA.91.022505}}.

\bibitem{Bilenky2012}
S.~M. Bilenky, C.~Giunti, Mod. Phys. Lett. A 27~(13) (2012) 1230015.
\newblock \href {http://dx.doi.org/10.1142/S0217732312300157}
  {\path{doi:10.1142/S0217732312300157}}.

\bibitem{Danilov2000}
M.~Danilov, R.~DeVoe, A.~Dolgolenko, {et al.}, Phys. Lett. B 480 (2000) 12.
\newblock \href {http://dx.doi.org/10.1016/S0370-2693(00)00404-4}
  {\path{doi:10.1016/S0370-2693(00)00404-4}}.

\bibitem{Sin11}
D.~Sinclair, E.~Rollin, J.~Smith, {et al.}, J. Phys.: Conf. Ser. 309 (2011)
  012005.
\newblock \href {http://dx.doi.org/10.1088/1742-6596/309/1/012005}
  {\path{doi:10.1088/1742-6596/309/1/012005}}.

\bibitem{Green2007}
M.~Green, J.~Wodin, R.~DeVoe, {et al.}, Phys. Rev. A 76 (2007) 023404.
\newblock \href {http://dx.doi.org/10.1103/PhysRevA.76.023404}
  {\path{doi:10.1103/PhysRevA.76.023404}}.

\bibitem{Diez2010}
M.~Montero~D\'iez, K.~Twelker, W.~Fairbank, {et al.}, Rev. of Sci. Instrum. 81
  (2010) 113301.
\newblock \href {http://dx.doi.org/10.1063/1.3499505}
  {\path{doi:10.1063/1.3499505}}.

\bibitem{NuDat}
{National Nuclear Data Center}, \href{www.nndc.bnl.gov/nudat2/}{information
  extracted from the nudat 2 database}.
\newline\urlprefix\url{www.nndc.bnl.gov/nudat2/}

\bibitem{Ziegler2010}
J.~F. Ziegler, M.~Ziegler, J.~Biersack, Nucl. Instrum. Meth. B 268 (2010) 1818.
\newblock \href {http://dx.doi.org/10.1016/j.nimb.2010.02.091}
  {\path{doi:10.1016/j.nimb.2010.02.091}}.

\bibitem{Borges1996}
F.~I. Borges, C.~Conde, {Nucl. Instrum. Meth. A} 381~(1) (1996) 91.
\newblock \href {http://dx.doi.org/10.1016/0168-9002(96)00739-5}
  {\path{doi:10.1016/0168-9002(96)00739-5}}.

\bibitem{Bru13}
T.~Brunner, D.~Fudenberg, A.~Sabourov, {et al.}, Nucl. Instrum. Meth. B 317,
  Part B (2013) 473.
\newblock \href {http://dx.doi.org/10.1016/j.nimb.2013.05.086}
  {\path{doi:10.1016/j.nimb.2013.05.086}}.

\bibitem{Maj68}
F.~G. Major, H.~G. Dehmelt, Phys. Rev. 170 (1968) 91.
\newblock \href {http://dx.doi.org/10.1103/PhysRev.170.91}
  {\path{doi:10.1103/PhysRev.170.91}}.

\bibitem{Schwarz08}
S.~Schwarz, Simulations for Ion Traps Buffer Gas Cooling. Trapped Charged
  Particles and Fundamental Interactions, Springer, Berlin, 2008.
\newblock \href {http://dx.doi.org/010.1007/978-3-540-77817-2 4}
  {\path{doi:010.1007/978-3-540-77817-2 4}}.

\bibitem{DeVoe09}
R.~G. DeVoe, Phys. Rev. Lett. 102 (2009) 063001.
\newblock \href {http://dx.doi.org/10.1103/PhysRevLett.102.063001}
  {\path{doi:10.1103/PhysRevLett.102.063001}}.

\bibitem{Maj05}
F.~Major, V.~Gheorghe, G.~Werth, Charged Particle Traps. Physics and Techniques
  of Charged Particle Field Confinement, Springer, Berlin, 2005.

\bibitem{Gerlich1992}
D.~Gerlich, {Adv. Chem. Phys.} {82} ({1992}) {1}.
\newblock \href {http://dx.doi.org/10.1002/9780470141397.ch1}
  {\path{doi:10.1002/9780470141397.ch1}}.

\bibitem{Dahl00}
D.~A. Dahl, Int. J. Mass Spectrom. 200 (2000) 3.
\newblock \href {http://dx.doi.org/10.1016/S1387-3806(00)00305-5}
  {\path{doi:10.1016/S1387-3806(00)00305-5}}.

\bibitem{Konenkov2010}
A.~Konenkov, D.~Douglas, N.~Konenkov, Int. J. Mass Spectrom. 289 (2010) 144.
\newblock \href {http://dx.doi.org/10.1016/j.ijms.2009.10.007}
  {\path{doi:10.1016/j.ijms.2009.10.007}}.

\bibitem{Varentsov1998}
V.~Varentsov, A.~Ignatiev, {Nucl. Instrum. Meth. A} 413 (1998) 447.
\newblock \href {http://dx.doi.org/10.1016/S0168-9002(98)00354-4}
  {\path{doi:10.1016/S0168-9002(98)00354-4}}.

\bibitem{EdwardsPump}
D.~Steele, {Edwards, private communication} (Sept 2014).

\bibitem{Schwartz2002}
J.~C. Schwartz, M.~W. Senko, J.~E. Syka, J. Am. Soc. Mass Spectrom. 13~(6)
  (2002) 659.
\newblock \href {http://dx.doi.org/10.1016/S1044-0305(02)00384-7}
  {\path{doi:10.1016/S1044-0305(02)00384-7}}.

\bibitem{Varentsov2008}
V.~L. Varentsov, Proc. SPIE 7025 (2008) 702509--702509--12.
\newblock \href {http://dx.doi.org/10.1117/12.802356}
  {\path{doi:10.1117/12.802356}}.

\bibitem{Dieteker2012}
R.~Dietiker, T.~Egorova, D.~G{\"u}nther, B.~Hattendorf, V.~Varentsov, {WO
  Patent App. PCT/EP2011/003,256} (Jan.~12 2012).
\newblock \href{www.google.com/patents/WO2012003946A1?cl=en}{[link]}.
\newline\urlprefix\url{www.google.com/patents/WO2012003946A1?cl=en}

\bibitem{Chen1999}
C.~Y. Chen, Y.~M. Li, K.~Bailey, T.~P. O'Connor, {et al.}, Science 286 (1999)
  1139.
\newblock \href {http://dx.doi.org/10.1126/science.286.5442.1139}
  {\path{doi:10.1126/science.286.5442.1139}}.

\bibitem{Jiang2011}
W.~Jiang, W.~Williams, K.~Bailey, A.~M. Davis, {et al.}, Phys. Rev. Lett. 106
  (2011) 103001.
\newblock \href {http://dx.doi.org/10.1103/PhysRevLett.106.103001}
  {\path{doi:10.1103/PhysRevLett.106.103001}}.

\bibitem{Dobi2012}
A.~Dobi, C.~Hall, S.~Slutsky, Y.-R. Yen, {et al.}, {Nucl. Instrum. Meth. A} 675
  (2012) 40.
\newblock \href {http://dx.doi.org/10.1016/j.nima.2012.01.066}
  {\path{doi:10.1016/j.nima.2012.01.066}}.

\bibitem{Ranjan2011}
M.~Ranjan, S.~Purushothaman, T.~Dickel, H.~Geissel, {et al.}, Europhys. Lett.
  96 (2011) 52001.
\newblock \href {http://dx.doi.org/10.1209/0295-5075/96/52001}
  {\path{doi:10.1209/0295-5075/96/52001}}.

\end{thebibliography}
\bibliographystyle{elsarticle-num}
\end{document}